\RequirePackage{lineno}
\documentclass[prd,twocolumn,showpacs,preprintnumbers,nofootinbib,amsmath,amssymb,nobalancelastpage,superscriptaddress]{revtex4-1}

\usepackage{graphicx}
\usepackage{bm}
\usepackage{dcolumn}
\usepackage{amsmath}
\usepackage{amssymb}
\usepackage{color,soul}
\usepackage{aas_macros}
\usepackage{url}
\usepackage{hyperref}
\usepackage{hyperref}

\newcommand{\cm}{{\mathrm{cm}}}
\newcommand{\second}{{\mathrm{s}}}

\newcommand{\Emin}{E_{\mathrm{min}}}

\newcommand{\sA}{\sigma}
\newcommand{\sigv}{\langle \sA v \rangle}

\newcommand{\Aeff}{A_\mathrm{eff}}
\newcommand{\Tobs}{T_\mathrm{obs}}
\newcommand{\prob}{\mathrm{P}}

\newcommand{\PSF}{{\rm PSF}}

\newcommand\nhat{\hat{\mathbf{n}}}

\newcommand\CLs{{\rm CL}_s}

\begin{document}
\title{Dark Matter Constraints from a Joint Analysis of Dwarf Spheroidal Galaxy Observations with VERITAS}

\author{S.~Archambault}
\affiliation{Physics Department, McGill University, Montreal, QC H3A 2T8, Canada}
\author{A.~Archer}
\affiliation{Department of Physics, Washington University, St. Louis, MO 63130, USA}
\author{W.~Benbow}
\affiliation{Fred Lawrence Whipple Observatory, Harvard-Smithsonian Center for Astrophysics, Amado, AZ 85645, USA}
\author{R.~Bird}
\affiliation{Department of Physics and Astronomy, University of California, Los Angeles, CA 90095, USA}
\author{E.~Bourbeau}
\affiliation{Physics Department, McGill University, Montreal, QC H3A 2T8, Canada}
\author{T.~Brantseg}
\affiliation{Department of Physics and Astronomy, Iowa State University, Ames, IA 50011, USA}
\author{M.~Buchovecky}
\affiliation{Department of Physics and Astronomy, University of California, Los Angeles, CA 90095, USA}
\author{J.~H.~Buckley}
\affiliation{Department of Physics, Washington University, St. Louis, MO 63130, USA}
\author{V.~Bugaev}
\affiliation{Department of Physics, Washington University, St. Louis, MO 63130, USA}
\author{K.~Byrum}
\affiliation{Argonne National Laboratory, 9700 S. Cass Avenue, Argonne, IL 60439, USA}
\author{M.~Cerruti}
\affiliation{Fred Lawrence Whipple Observatory, Harvard-Smithsonian Center for Astrophysics, Amado, AZ 85645, USA}
\author{J.~L.~Christiansen}
\affiliation{Physics Department, California Polytechnic State University, San Luis Obispo, CA 94307, USA}
\author{M.~P.~Connolly}
\affiliation{School of Physics, National University of Ireland Galway, University Road, Galway, Ireland}
\author{W.~Cui}
\affiliation{Department of Physics and Astronomy, Purdue University, West Lafayette, IN 47907, USA}
\affiliation{Department of Physics and Center for Astrophysics, Tsinghua University, Beijing 100084, China.}
\author{M.~K.~Daniel}
\affiliation{Fred Lawrence Whipple Observatory, Harvard-Smithsonian Center for Astrophysics, Amado, AZ 85645, USA}
\author{Q.~Feng}
\affiliation{Physics Department, McGill University, Montreal, QC H3A 2T8, Canada}
\author{J.~P.~Finley}
\affiliation{Department of Physics and Astronomy, Purdue University, West Lafayette, IN 47907, USA}
\author{H.~Fleischhack}
\affiliation{DESY, Platanenallee 6, 15738 Zeuthen, Germany}
\author{L.~Fortson}
\affiliation{School of Physics and Astronomy, University of Minnesota, Minneapolis, MN 55455, USA}
\author{A.~Furniss}
\affiliation{Department of Physics, California State University - East Bay, Hayward, CA 94542, USA}
\author{A.~Geringer-Sameth}
\email{a.geringer-sameth@imperial.ac.uk}
\affiliation{Department of Physics, Brown University, Providence, RI 02912, USA}
\affiliation{McWilliams Center for Cosmology, Department of Physics, Carnegie Mellon University, Pittsburgh, Pennsylvania 15213, USA}
\author{S.~Griffin}
\affiliation{Physics Department, McGill University, Montreal, QC H3A 2T8, Canada}
\author{J.~Grube}
\affiliation{Department of Physics, Stevens Institute of Technology, Hoboken, NJ 07030, USA}
\author{M.~H{\"u}tten}
\affiliation{DESY, Platanenallee 6, 15738 Zeuthen, Germany}
\author{N.~H{\aa}kansson}
\affiliation{Institute of Physics and Astronomy, University of Potsdam, 14476 Potsdam-Golm, Germany}
\author{D.~Hanna}
\affiliation{Physics Department, McGill University, Montreal, QC H3A 2T8, Canada}
\author{O.~Hervet}
\affiliation{Santa Cruz Institute for Particle Physics and Department of Physics, University of California, Santa Cruz, CA 95064, USA}
\author{J.~Holder}
\affiliation{Department of Physics and Astronomy and the Bartol Research Institute, University of Delaware, Newark, DE 19716, USA}
\author{G.~Hughes}
\affiliation{Fred Lawrence Whipple Observatory, Harvard-Smithsonian Center for Astrophysics, Amado, AZ 85645, USA}
\author{B.~Hummensky}
\affiliation{Physics Department, Columbia University, New York, NY 10027, USA}
\author{C.~A.~Johnson}
\affiliation{Santa Cruz Institute for Particle Physics and Department of Physics, University of California, Santa Cruz, CA 95064, USA}
\author{P.~Kaaret}
\affiliation{Department of Physics and Astronomy, University of Iowa, Van Allen Hall, Iowa City, IA 52242, USA}
\author{P.~Kar}
\affiliation{Department of Physics and Astronomy, University of Utah, Salt Lake City, UT 84112, USA}
\author{N.~Kelley-Hoskins}
\affiliation{DESY, Platanenallee 6, 15738 Zeuthen, Germany}
\author{M.~Kertzman}
\affiliation{Department of Physics and Astronomy, DePauw University, Greencastle, IN 46135-0037, USA}
\author{D.~Kieda}
\affiliation{Department of Physics and Astronomy, University of Utah, Salt Lake City, UT 84112, USA}
\author{S.~Koushiappas}
\email{koushiappas@brown.edu}
\affiliation{Department of Physics, Brown University, Providence, RI 02912, USA}
\affiliation{Institute for Theory and Computation, Harvard University, 60 Garden St., Cambridge, MA 02138}
\author{M.~Krause}
\affiliation{DESY, Platanenallee 6, 15738 Zeuthen, Germany}
\author{F.~Krennrich}
\affiliation{Department of Physics and Astronomy, Iowa State University, Ames, IA 50011, USA}
\author{M.~J.~Lang}
\affiliation{School of Physics, National University of Ireland Galway, University Road, Galway, Ireland}
\author{T.~T.Y.~Lin}
\affiliation{Physics Department, McGill University, Montreal, QC H3A 2T8, Canada}
\author{S.~McArthur}
\affiliation{Department of Physics and Astronomy, Purdue University, West Lafayette, IN 47907, USA}
\author{P.~Moriarty}
\affiliation{School of Physics, National University of Ireland Galway, University Road, Galway, Ireland}
\author{R.~Mukherjee}
\affiliation{Department of Physics and Astronomy, Barnard College, Columbia University, NY 10027, USA}
\author{D.~Nieto}
\affiliation{Physics Department, Columbia University, New York, NY 10027, USA}
\author{S.~O'Brien}
\affiliation{School of Physics, University College Dublin, Belfield, Dublin 4, Ireland}
\author{R.~A.~Ong}
\affiliation{Department of Physics and Astronomy, University of California, Los Angeles, CA 90095, USA}
\author{A.~N.~Otte}
\affiliation{School of Physics and Center for Relativistic Astrophysics, Georgia Institute of Technology, 837 State Street NW, Atlanta, GA 30332-0430}
\author{N.~Park}
\affiliation{Enrico Fermi Institute, University of Chicago, Chicago, IL 60637, USA}
\author{M.~Pohl}
\affiliation{Institute of Physics and Astronomy, University of Potsdam, 14476 Potsdam-Golm, Germany}
\affiliation{DESY, Platanenallee 6, 15738 Zeuthen, Germany}
\author{A.~Popkow}
\affiliation{Department of Physics and Astronomy, University of California, Los Angeles, CA 90095, USA}
\author{E.~Pueschel}
\affiliation{School of Physics, University College Dublin, Belfield, Dublin 4, Ireland}
\author{J.~Quinn}
\affiliation{School of Physics, University College Dublin, Belfield, Dublin 4, Ireland}
\author{K.~Ragan}
\affiliation{Physics Department, McGill University, Montreal, QC H3A 2T8, Canada}
\author{P.~T.~Reynolds}
\affiliation{Department of Physical Sciences, Cork Institute of Technology, Bishopstown, Cork, Ireland}
\author{G.~T.~Richards}
\affiliation{School of Physics and Center for Relativistic Astrophysics, Georgia Institute of Technology, 837 State Street NW, Atlanta, GA 30332-0430}
\author{E.~Roache}
\affiliation{Fred Lawrence Whipple Observatory, Harvard-Smithsonian Center for Astrophysics, Amado, AZ 85645, USA}
\author{C.~Rulten}
\affiliation{School of Physics and Astronomy, University of Minnesota, Minneapolis, MN 55455, USA}
\author{I.~Sadeh}
\affiliation{DESY, Platanenallee 6, 15738 Zeuthen, Germany}
\author{M.~Santander}
\affiliation{Department of Physics and Astronomy, Barnard College, Columbia University, NY 10027, USA}
\author{G.~H.~Sembroski}
\affiliation{Department of Physics and Astronomy, Purdue University, West Lafayette, IN 47907, USA}
\author{K.~Shahinyan}
\affiliation{School of Physics and Astronomy, University of Minnesota, Minneapolis, MN 55455, USA}
\author{A.~W.~Smith}
\affiliation{University of Maryland, College Park / NASA GSFC, College Park, MD 20742, USA}
\author{D.~Staszak}
\affiliation{Enrico Fermi Institute, University of Chicago, Chicago, IL 60637, USA}
\author{I.~Telezhinsky}
\affiliation{Institute of Physics and Astronomy, University of Potsdam, 14476 Potsdam-Golm, Germany}
\affiliation{DESY, Platanenallee 6, 15738 Zeuthen, Germany}
\author{S.~Trepanier}
\affiliation{Physics Department, McGill University, Montreal, QC H3A 2T8, Canada}
\author{J.~V.~Tucci}
\affiliation{Department of Physics and Astronomy, Purdue University, West Lafayette, IN 47907, USA}
\author{J.~Tyler}
\affiliation{Physics Department, McGill University, Montreal, QC H3A 2T8, Canada}
\author{S.~P.~Wakely}
\affiliation{Enrico Fermi Institute, University of Chicago, Chicago, IL 60637, USA}
\author{A.~Weinstein}
\affiliation{Department of Physics and Astronomy, Iowa State University, Ames, IA 50011, USA}
\author{P.~Wilcox}
\affiliation{Department of Physics and Astronomy, University of Iowa, Van Allen Hall, Iowa City, IA 52242, USA}
\author{D.~A.~Williams}
\affiliation{Santa Cruz Institute for Particle Physics and Department of Physics, University of California, Santa Cruz, CA 95064, USA}
\author{B.~Zitzer}
\email{bzitzer@physics.mcgill.ca}
\affiliation{Physics Department, McGill University, Montreal, QC H3A 2T8, Canada}

\collaboration{The VERITAS Collaboration}


\date{\today}

\begin{abstract}
We present constraints on the annihilation cross section of WIMP dark matter based on the joint statistical analysis of four dwarf galaxies with VERITAS. These results are derived from an optimized photon weighting statistical technique that improves on standard imaging atmospheric Cherenkov telescope (IACT) analyses by utilizing the spectral and spatial properties of individual photon events. We report on the results of $\sim$230 hours of observations of five dwarf galaxies and the joint statistical analysis of four of the dwarf galaxies. We find no evidence of gamma-ray emission from any individual dwarf nor in the joint analysis. The derived upper limit on the dark matter annihilation cross section from the joint analysis is $1.35\times 10^{-23} {\mathrm{ cm^3s^{-1}}}$ at 1 TeV for the bottom quark ($b\bar{b}$) final state, $2.85\times 10^{-24}{\mathrm{ cm^3s^{-1}}}$ at 1 TeV for the tau lepton ($\tau^{+}\tau^{-}$) final state and  $1.32\times 10^{-25}{\mathrm{ cm^3s^{-1}}}$ at 1 TeV for the gauge boson ($\gamma\gamma$) final state. 
\end{abstract}


\maketitle

\section{Introduction}
The search for standard model particles resulting from the annihilation of dark matter particles provides an important complement to the efforts of direct searches for dark matter interactions and searches for dark matter production at particle accelerators. Among the theoretical candidates for the dark matter particle above a few GeV, Weakly Interacting Massive Particles (WIMPs) are well motivated \cite{1996PhR...267..195J,2003NuPhB.650..391S} as they naturally provide the measured present day cold dark matter density \cite{2014A&A...571A..16P,1979ARNPS..29..313S,1965PhL....17..164Z,zel1965advances,1966PhRvL..17..712C}.  In such models, the WIMPs either decay or annihilate into standard model particles that produce mono-energetic gamma-ray lines and/or a continuum of gamma rays with energies up to the dark matter particle mass.

Attractive targets for indirect dark matter searches are nearby massive objects with high inferred dark matter content and that are not expected to be sources of very-high-energy gamma rays.  Dwarf spheroidal galaxies (dSphs) are relatively close ($\sim$20 to 200 kpc) to Earth and lack conventional astrophysical high-energy sources of gamma rays \cite{1998ARA&A..36..435M}. Five dwarf galaxies have been observed with the Very Energetic Radiation Imaging Telescope Array System (VERITAS) between 2007 and 2013, for a total of 230 hours of high quality data.

In this paper we perform a joint statistical analysis of dwarf galaxies observed with VERITAS. We find no evidence of dark matter annihilation in any of the dwarf galaxies individually observed with VERITAS or in a joint analysis of four of the dwarfs. We place upper limits on the emitted flux and derive upper limits on the annihilation cross section.

\section{Observations} 

VERITAS  is an array of four imaging atmospheric Cherenkov telescopes (IACTs), each 12 m in diameter, located at the Fred Lawrence Whipple Observatory in southern Arizona, USA (31.68$^{\circ}$ N, 110.95$^{\circ}$ W, 1.3 km above sea level). Each VERITAS camera contains 499 pixels (0.15$^{\circ}$ diameter) and has a field of view of 3.5$^{\circ}$. 
VERITAS began full array operations in the spring of 2007. The instrument has gone through a number of upgrades since then to improve performance.  In the summer of 2009, the first telescope (``T1'') was moved to its current location in the array to provide a more uniform distance between telescopes, improving the sensitivity of the system \cite{2009arXiv0912.3841P}. The telescope-level trigger was replaced with a faster system in the fall of 2011 \cite{2013arXiv1307.8360Z}, allowing for greater night-sky background (NSB) reduction during all operating modes of the experiment. The VERITAS camera pixels were replaced in summer 2012 with higher quantum efficiency photomultiplier tubes (PMTs), allowing for a lowered energy threshold \cite{2013arXiv1308.4849D}. VERITAS is sensitive to gamma rays from approximately 85 GeV (after camera upgrade) to greater than 30 TeV with a typical energy resolution of $15-25\%$  and an angular resolution (68\% containment) of $<$0.1 degrees per event.  The flux sensitivity of the standard analysis is such that a source with a flux of order of 1\% of the Crab Nebula flux can be detected in approximately 25 hours of observation. The looser event selection criteria (commonly referred to as ``cuts") used in this work described later in this section resulted in a slightly larger energy resolution (25\%-30\% at 1 TeV) and angular resolution ($\sim$0.12$^{\circ}$ at 1 TeV).

From the beginning of four-telescope operations in 2007 to the summer of 2013, five dwarf galaxies in the northern hemisphere have been observed by VERITAS: Segue 1, Ursa Minor, Draco, Bo\"{o}tes and Willman 1. Quality data for this analysis requires moonless and clear atmospheric (based on infrared temperature measurements) conditions and operation of all four telescopes. Dwarf galaxy data used in this work were taken during three different epochs of VERITAS operations: data taken before the move of T1, data taken after the T1 move, and data taken after the camera upgrade.  
In all three epochs, data were obtained with the wobble pointing strategy, where the camera center is offset by 0.5 degrees from the target position \cite{1994APh.....2..137F}. The wobble mode allows for simultaneous background estimation and source observation, reducing the systematic uncertainties in the background estimation as opposed to using separate pointings for background estimation. 

The data reduction mostly follows the standard techniques employed by VERITAS~\cite{2006APh....25..391H}, with the notable exceptions being the methodology of the cosmic-ray background estimate, the adopted statistical approach based on individual photon weighting, and the method of image characterization for shower reconstruction.  Images recorded by the VERITAS cameras are calibrated by the photomultiplier tube (PMT) gains. Traditionally the showers are characterized by their second moments~\cite{1985ICRC....3..445H}. In this work each Cherenkov shower image is fit with a two-dimensional elliptical Gaussian function to get the parameter characterization of the shower  \cite{2012AIPC.1505..709C}.  This fitting method for Cherenkov images is advantageous because the two-dimensional elliptical Gaussian fit allows for better point-spread function (PSF) characterization at high energies, and is less biased to images that are truncated at the edge of the camera or by dead pixels or suppressed pixels due to bright stars. This method of fitting has also been shown to reduce the time for a weak point source to reach 5$\sigma$ by 20\% \cite{2012AIPC.1505..709C}. The stereo reconstruction of the event's arrival direction and energy is accomplished by combining parameters from multiple telescopes \cite{2006APh....25..380K}. The hadronic cosmic-ray background is reduced by applying mean scaled width and mean scaled length cuts \cite{2006APh....25..380K}. The cuts were optimized {\it a priori} using data from known weak and soft-spectral very-high-energy sources. These ``soft" cuts were selected to give the lowest possible energy threshold, which increases sensitivity to dark matter searches by allowing more low energy events to be used for the analysis. An additional cut is applied on the angle between the target position and the reconstructed arrival position, $\theta < 0.17 $ degrees, thus defining the signal search region or ``ON region''. 

Many IACT analyses select background events from one or more OFF regions in the camera field of view \cite{2007A&A...466.1219B}. Two methods for forming an OFF region are commonly used. In the reflected region method (also called a wobble analysis), the source is offset from the telescope tracking position, and OFF regions consist of regions with the same size as the ON region with the same offset. In the ring background method the OFF region is an annulus surrounding the ON region.

\begin{figure}
	\centering
	\includegraphics[clip,trim={0 6cm 0 0}, scale=0.35]{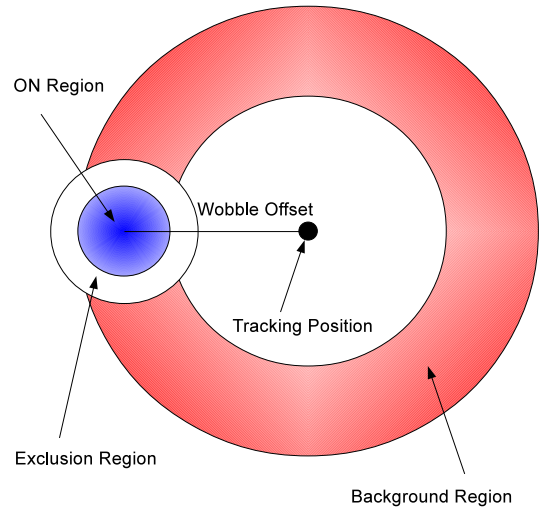}
	\caption{Illustration of the background method that is used for the photon weighting analysis in the dark matter search. The ON region is shaded in light blue, while the OFF region is shaded in red. Note that this figure is not drawn to scale. The standard offset from the center of the ON region to the tracking position is 0.5$^{\circ}$.}
	\label{fig:CBM}
\end{figure}

This analysis requires a larger sample of the measured background and to determine its energy spectrum, therefore a third method is introduced. We name this new method the ``crescent'' background method (CBM) \cite{2013arXiv1307.8367Z}. This method was previously described in  \citet{2007A&A...466.1219B} but this is the first time it has been applied to IACT data. Background events are selected from an annulus similar to the ring background. However, the annulus is centered on  the {\it tracking} position 
as opposed to the {\it source} position (see Figure~\ref{fig:CBM}). This gives roughly a factor of two more background events than from standard reflected regions (depending on the field of view of the array pointing). The ring background method typically used is not suitable for this analysis, due to the energy dependence in IACT acceptances. Those acceptances are symmetrical around the tracking position to first order \cite{2007A&A...466.1219B}. By selecting events only from a region at approximately the same angular distance from the tracking position, we reduce the energy dependence of the background scaling factor, $\alpha$.

Visible starlight may bias the background estimate and is removed by defining circular background exclusion regions centered around stars with apparent magnitudes of $m_{\mathrm{B}} < 8$. The size of the exclusion region used varies with the brightness of the star; for example an exclusion region of 0.4 degrees is set around the 3.5-apparent magnitude star $\eta$ Leonis in the field of Segue 1. The central region of radius 0.3 degrees around each dwarf is also excluded. 

The scaling factor of each background event, $\alpha$, used to calculate the gamma-ray excess and significance \cite{1983ApJ...272..317L} is determined by the ratio of the integral of the cosmic-ray acceptance within the ON region to the integral of the acceptance within the crescent-shaped OFF region. To better account for background systematics associated with deep exposures, an acceptance function was derived using the zenith angle of observation as well as the angular distance from the tracking direction. The procedure is similiar to the one described in the appendix of \citet{2003A&A...410..389R} and is described in more detail in \cite{2015arXiv150300743B}.  An acceptance gradient in the VERITAS cameras was determined by utilizing a smoothed map of the ratio of counts using the total data set for each dSph in each skymap bin to the azimuthally-symmetric acceptance in that map bin, a parameter we refer to as {\it flatness}. If the radial-only acceptance adequately describes the cosmic-ray background, then the flatness map should be uniform within statistical errors across the field of view, i.e. it should not correlate with zenith angle or any other external parameters. A second map was produced with the mean difference of the zenith angle of the reconstructed photon direction from the zenith angle of the array tracking direction at the time the event was recorded. We will refer to this  as the mean zenith map for simplicity. A scatter plot of the contents of each bin for the mean zenith map and the flatness map was made, showing a strong correlation for each field of view. That correlation was fit with a fourth-degree polynomial which was used to re-weight each bin in the spatial acceptance map and re-calculate $\alpha$. The difference between $\alpha$ with and without the zenith correction is $\lesssim$ 1\%.

\begin{figure*}[t]
	\centering
	\includegraphics[width=0.4\textwidth]{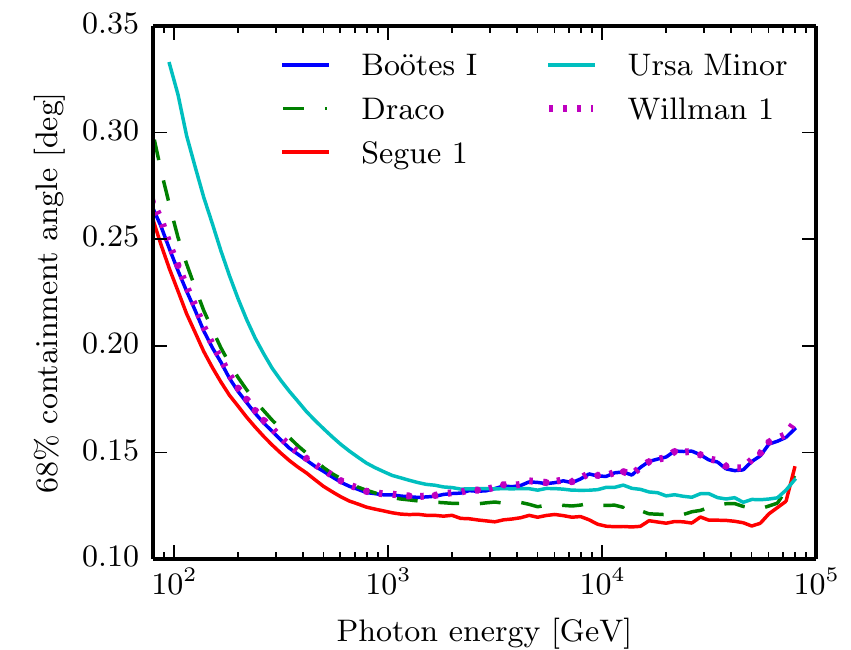}
	\includegraphics[width=0.4\textwidth]{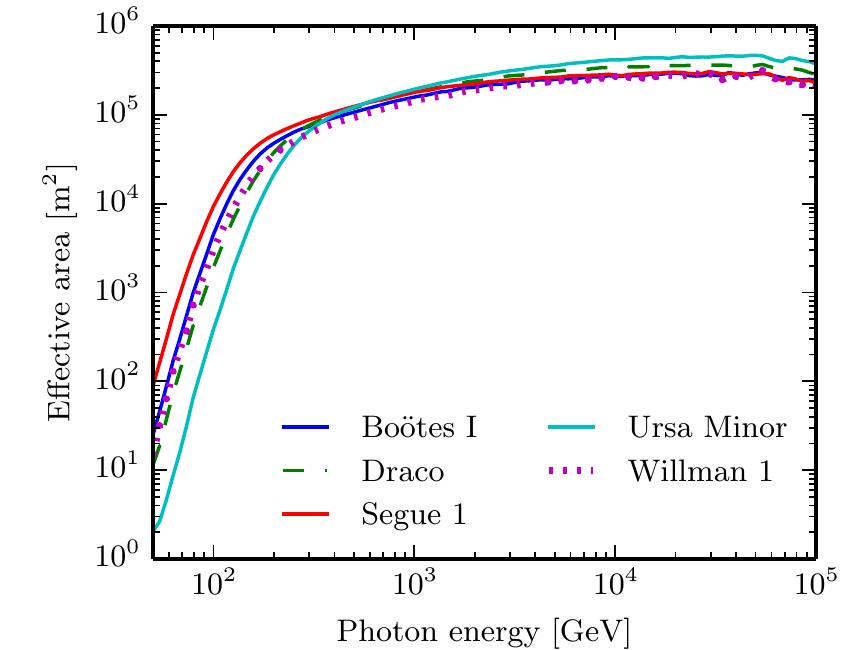}
	\caption{Mean point-spread function (left panel) and mean effective areas (right panel) vs. Monte Carlo (MC) energy for the observing conditions of the five dwarf spheroidals in this work. }
	\label{fig:response}
\end{figure*}

\section{Dark Matter Distribution within the dwarfs \label{sec:Jprofiles}}

The strength of the predicted gamma-ray signal is proportional to the dark matter distribution within dwarf galaxies. In general, this is characterized by the $J$-profile, defined as 
\begin{eqnarray}
\frac{dJ(\nhat)}{d\Omega} = \int \,  \rho^2(\ell \nhat) \, d\ell ,
\end{eqnarray}
where $\ell$ is the line-of-sight distance along the  $\nhat$ direction, $d\Omega$ is the solid angle, and $\rho$ is the mass density profile of the dwarf galaxy. 

The distribution of dark matter in dwarf galaxies is obtained using line-of-sight velocity and position measurements of stars that are gravitationally bound within the dwarf galaxy potential well~\cite{simon07,walker09a}. Distributions of stellar velocities and positions are functions of the gravitational potential as described by the Jeans equation~\cite{2007PhRvD..75h3526S,2008ApJ...678..614S,2008gady.book.....B,2013pss5.book.1039W,2013NewAR..57...52B,2013PhR...531....1S}.

We adopt the observational constraints on $J$-profiles as derived by~ \citet{2015ApJ...801...74G}. The density profile of each dwarf is modeled as a ``generalized'' NFW (Navarro-Frenk-White) profile  \cite{1996MNRAS.278..488Z}, 
\begin{eqnarray}
\rho(r) = \rho_s [r/r_s]^{-\gamma} [1+(r/r_s)^\alpha]^{(\gamma-\beta)/\alpha}, 
\end{eqnarray}
with five free parameters. A likelihood function relates the five parameters (and a sixth nuisance parameter specifying the stellar velocity anisotropy) to the observables through the Jeans equation. The parameter space is explored, giving rise to a chain of posterior sample halos.

This analysis generates many realizations of halos which reasonably fit the stellar kinematic data. This produces a systematic uncertainty for the dark matter search. When we present the results of the search and limits on the annihilation cross section we will separate this systematic uncertainty from the statistical uncertainty induced by our finite event statistics. This is done by repeating the analysis separately for different realizations of halo parameters. The systematic uncertainty ``band" that results from this repetition should be thought of as reflecting our imperfect knowledge of the dwarf density profiles. See Section IX.C of ~\cite{2015PhRvD..91h3535G} for details.

Use of the Jeans equation requires the assumption that stellar tracers are in dynamical equilibrium and the analysis of \cite{2015ApJ...801...74G} further assumes spherical symmetry, Plummer light profiles, and velocity anisotropy that is constant with radius. These are approximations, and all real systems will violate them at some level. Bonnivard et. al. \cite{2015MNRAS.446.3002B} have studied the biases introduced by these effects. While the statistical uncertainty due to finite kinematic sample sizes dominates the errors in $J$ for ultrafaint dwarfs (e.g. Segue 1, Bo{\"o}tes 1, Willman 1), the assumption of spherical symmetry may cause a moderate bias (comparable to the statistical error bar) for the classical dwarfs (e.g. Draco, Ursa Minor). In the combined analysis, the uncertainties for Segue 1 dominate the error budget and our results will be insensitive to the other systematic effects mentioned above.

The stellar population of Willman 1 shows irregular kinematics, which may be due to ongoing tidal disruption of the satellite \cite{2011AJ....142..128W}. Regardless of the cause, the observations strongly suggest that Willman 1 is not in dynamical equilibrium, violating a core assumption of the Jeans equation. This object was excluded from the analysis of  \citet{2015ApJ...801...74G}, who considered the inferred $J$-profile to be unreliable with no handle on the magnitude of the error. In the present work, we therefore exclude Willman 1 from results which require an estimate of its $J$-profile.

Additionally, Bonnivard et. al. \cite{2016MNRAS.462..223B} have pointed out the possibility of contamination of the stellar samples used to perform the Jeans analysis. Milky Way interlopers mistakenly included in the spectroscopic sample of dwarf member stars will inflate the inferred velocity dispersion and may bias $J$-profiles toward large expected annihilation signals. In particular, there are indications that Segue 1 may suffer from such contamination: the removal of several ambiguous stars from Segue 1 sample can have drastic (i.e. orders of magnitude) effects on $J$.  Compared with classical dwarfs, this issue will be most severe for ultrafaint dwarfs, which have much smaller spectroscopic samples. While several groups have begun extending the Jeans analysis framework to encompass foreground contamination\cite{2016MNRAS.462..223B}\cite{2016arXiv160801749I}\cite{2016MNRAS.463.1117Z}, no uniform analysis of the dwarf population has been performed, though several groups have begun extending the analysis framework to encompass this effect \cite{2016arXiv160801749I}\cite{2016MNRAS.463.1117Z}. Notably, the issue of contamination has not been observationally checked for any ultrafaint dwarfs apart from Segue~1 and the recently discovered Reticulum~II. Ichikawa et. al. \cite{2016arXiv160801749I}, simulating future spectroscopic observations, find that contamination may bias $J$ high by factors of $\sim 3$ for the classical dwarfs Draco and Ursa Minor. Therefore, we caution that the uncertainties in our particle physics limits may be underestimated due to this additional astrophysical systematic uncertainty.

\section{Event weighting }
We employ a newly-developed event weighting method~\cite{2015PhRvD..91h3535G} to simultaneously analyze the data from all five dwarf fields. This technique improves on standard IACT analyses by utilizing the spectral and spatial properties of the individual events. It also takes into account the expected properties of the annihilation signal and the instrumental and astrophysical backgrounds, to perform an ``optimal'' analysis (see \cite{2015PhRvD..91h3535G} for further details and a theoretical development of the technique). 

Given the reconstructed events in an ON region we seek an optimal way to extract a possible dark matter signal. Each reconstructed event is assigned a weight based on three parameters: the dwarf field $\nu$ it came from, its reconstructed energy $E$, and its reconstructed angular separation from the dwarf galaxy $\theta$. The test statistic $T$ is defined as 
\begin{eqnarray}
T = \sum\limits_i w_i,
\label{eqn:Tdef}
\end{eqnarray}
where the index $i$ runs over all ON events from all dwarf fields and $w_i = w(\nu_i, E_i, \theta_i)$ is the weight of the $i$th event.

The weight function $w(\nu, E, \theta)$ can be an arbitrary function of the event properties. For example, a conventional ON/OFF analysis (see e.g. \cite{2012PhRvD..85f2001A}) is recovered if $w=1$ for all events within the ON region of a particular dwarf and $w=0$ for all other events. In this case the test statistic is just the number of observed events in the ON region.

The weight function can be designed to distinguish, as efficiently as possible, the difference between background and background plus a dark matter signal. An intuitive solution is to weight different events according to how likely they are to be due to dark matter compared to background.

It has been shown \cite{2015PhRvD..91h3535G} that when testing a simple null hypothesis (background only) against a simple alternative (signal plus background) the optimal form of the weight function $w(\nu, E, \theta)$ is
\begin{eqnarray}
w = \log\left[ 1 + \frac{s}{b} \right],
\label{eqn:wdef}
\end{eqnarray}
where $s(\nu, E, \theta)$ is the expected number of signal events with properties $(\nu, E, \theta)$, and $b(\nu, E, \theta)$ is the expected number of background events due to all other processes besides dark matter annihilation (e.g. hadronic air showers, leptonic air showers and diffuse astrophysical gamma rays).  The test statistic derived from this weighting is optimal in the sense that it maximizes the statistical power of the hypothesis test; if a dark matter signal is hidden in the data this test statistic is most likely to turn up a detection (see~\cite{2015PhRvD..91h3535G} for details).

The functions $s(\nu, E, \theta)$ and $b(\nu, E, \theta)$ are differential quantities, namely the expected number of events from dwarf $\nu$ with energies between $E$ and $E+dE$ and angular separations between $\theta$ and $\theta + d\theta$. We use the events in the OFF region of each dwarf to estimate the function $b$. The energy spectrum of these background events is modeled as a piecewise function. For energies below 1 TeV we replace each event with a Gaussian of width 3\% of the measured energy, giving a kernel density estimate. This is a requirement of the kernel estimator and is unrelated to the VERITAS energy dispersion. Above 1 TeV we splice on a power law with exponential cutoff. The form is $f(E) =f_0(E/E_0)^\gamma \exp[(E-E_0)/E_{\mathrm{cut}} ]$, where $E_0 = \text{1 TeV}$ and $f_0$ is the kernel density estimate of the spectrum at 1 TeV. The choice of 3\% of the measured energy as well as 1 TeV for the energy cutoff are arbitrary and do not affect the statistical significances of the search or the coverage of the limits. The parameters $\gamma$ and $E_{\mathrm{cut}}$ are obtained using the unbinned maximum likelihood. We choose this smooth fitting function to avoid noise in the kernel density estimator due to the relatively low number of observed events with high energies. The corrected solid angle ratios $\alpha$  between OFF and ON regions are used to predict the expected number of background events in the ON region for each dwarf. The background is assumed to be isotropic within the ON region so the $\theta$ dependence of $b(\nu, E, \theta)$ is proportional to $\sin(\theta) d\theta$.

The expected signal $s(\nu, E, \theta)$ is determined by convolving the dark matter annihilation flux with the VERITAS instrument response. The gamma-ray flux from annihilation, i.e. flux of photons from direction $\nhat$ per energy per solid angle, is given by
\begin{eqnarray}
\frac{dF(E,\nhat)}{dEd\Omega} = \frac{\sigv}{8\pi M^2} \frac{dN_\gamma (E)}{dE} \frac{dJ(\nhat)}{d\Omega},
\label{eqn:dmflux}
\end{eqnarray}
where $M$ is the dark matter particle mass, $\sigv$ is the velocity-averaged annihilation cross section, and $dN_\gamma /dE$ is the spectrum of gamma rays from a single annihilation event. This last spectrum is determined by the branching ratios $B_i$ into the various Standard Model final states:
\begin{eqnarray}
\frac{dN_\gamma(E)}{dE} = \sum_i B_i \frac{dN_{\gamma,i}(E)}{dE},
\label{eqn:Bi}
\end{eqnarray}
where $dN_{\gamma,i}/dE$ is the number of gamma rays produced per annihilation per gamma-ray energy by the products of channel $i$. We adopt the annihilation spectra given in \cite{2011JCAP...03..051C}, including electroweak corrections. For annihilation into a two-photon final state we model the energy spectrum as a gaussian of width 10\% of the dark matter mass and an amplitude of two photons. This width is always less than the VERITAS energy resolution.  

The number of events reconstructed with energy $E$ and angular separation $\theta$ is given by the convolution
\begin{eqnarray}
\frac{dN(E,\nhat)}{dE d\Omega} = \int\limits_{E_t} \int\limits_{\Omega_t} dE_t d\Omega_t \frac{dF(E_t, \nhat_t)}{dE_t d\Omega_t} R(E, \nhat \vert E_t, \nhat_t),
\label{eqn:dNdEdO}
\end{eqnarray}
where the subscript $t$ denotes true energies and directions and the function $R$ is 
the response of VERITAS. For clarity we have omitted a subscript $\nu$ from the quantities in Eq.~\ref{eqn:dNdEdO}, but the predicted dark matter flux and VERITAS response depend on which dwarf is being considered.

The response $R(E, \nhat \vert E_t, \nhat_t) dE d\Omega$ is the probability (per incident flux) that a gamma ray with true energy $E_t$ and direction $\nhat_t$ will be reconstructed with an energy in the interval $dE$ around $E$ and in the solid angle $d\Omega$ around direction $\nhat$. It is the product (summed over VERITAS observation runs) of the effective area $\Aeff$, live time per observation run $\tau$, instrument PSF, and energy dispersion $D$:
\begin{eqnarray}
R(E, \nhat \vert E_t, \nhat_t) = \sum_\text{runs } \tau \Aeff(E_t)  \PSF(\nhat \vert E_t, \nhat_t) D(E \vert E_t).
\end{eqnarray}
These four factors are computed for each observation run. Because the considered $J$-profiles and PSFs are azimuthally symmetric in $\nhat$ (i.e. $dJ/d\Omega$ only depends on the angle between $\nhat$ and the dwarf and the PSF only depends on the angle between $\nhat$ and $\nhat_t$), the expected number of events is also azimuthally symmetric and depends only on $\theta$, the angle between the reconstructed direction $\nhat$ and the direction of the dwarf.

The VERITAS point spread function, $\PSF(\theta \vert E_{t})$ (probability per solid angle of detecting a photon of true energy $E_{t}$ an angular distance $\theta$ away from its true direction) is derived from gamma-ray simulations. The reason that simulations were used instead of data from a bright source (for example, the Crab Nebula) is that simulations provide much larger statistics, and therefore better characterization at all energies. The simulated PSF agrees well with Crab Nebula data, to within $\lesssim$10\% in the energy range where VERITAS is most sensitive. The same quality and background rejection cuts are applied to the simulated events, which are then binned in $\theta$ from 0$^{\circ}$ to 2$^{\circ}$ and in $E$ in the range from 0.01 TeV to 100 TeV, covering the entire VERITAS energy range. At each energy, the binned histogram is normalized over $\theta$, forming the probability distribution function, $\PSF(\theta \vert E_{t})$. The VERITAS epoch, the energy and the zenith angle are the only simulated parameters that have an impact on the shape of the PSF in this work, although others were investigated. Azimuthal angle and background noise dependencies have a negligible effect for this analysis. Examples of the energy dependence are shown in the left panel of Figure ~\ref{fig:response}. The differences in the curves are due to differences in zenith angle and the epochs the dSphs were observed in.

The effective collection area, $\Aeff(E_t)$ is a function of the true gamma-ray energy $E_t$, and it depends on the zenith and azimuth angles of observations, the amount of background noise present, VERITAS configuration epoch, offset of the source from the target position, and the gamma-ray cuts \cite{1998APh.....9...15M}. The right panel of Figure ~\ref{fig:response} depicts the average effective area curves of the observing conditions (zenith, azimuth, NSB and epochs) for all dwarf galaxies included in this study. 

The line spread function, or energy dispersion ($D(E \vert E_t)$) quantifies the energy resolution and bias of VERITAS. It is constructed by generating Monte Carlo gamma-ray showers at a true energy and putting the simulated showers through a simulated detector and the same reduction and cuts as the data. The shower reconstruction algorithm of the data analysis assigns the event a reconstructed energy $E$ \cite{1998APh.....9...15M}. Simulated showers that survive the ``soft'' cuts described above are put into a two dimensional histogram of reconstructed and true energy. Each bin of $E_t$ is normalized to unity to produce a probability density function.

Finally, the expected number of dark matter events from a dwarf with reconstructed energy between $E$ and $E+dE$ and separation between $\theta$ and $\theta + d\theta$ is simply
\begin{eqnarray}
s(\nu, E, \theta) = \frac{dN(\nu, E, \theta)}{dE d\Omega} dE \, 2\pi \sin(\theta) d\theta,
\label{eqn:s}
\end{eqnarray}
with $dN/dEd\Omega$ given by Eq.~\eqref{eqn:dNdEdO}.

%

To conduct a search for annihilation or set limits on the cross section we compute the probability distribution for measuring the test statistic under various hypotheses. For example, to conduct a search for dark matter annihilation, the observed value of the test statistic $\Tobs$ is compared with the probability distribution for $T$ due to background processes only $\prob(T \vert \text{bg-only})$. 
The significance of the detection is defined as the probability that $T$ is less than $\Tobs$ under the background-only hypothesis. It is convenient to convert this probability into a ``sigma value'' using percentiles of a standard Gaussian distribution.

Alternatively, to construct upper limits on the annihilation cross section we compute the distribution for $T$ given a particular dark matter model, which includes specifying values for the particle mass $M$, cross section $\sigv$, and the branching fractions $B_i$ (see Eqs.~\eqref{eqn:dmflux} and~\eqref{eqn:Bi}).

The method for computing the probability distribution for $T$ under any dark matter hypothesis (i.e. $\sigv \neq 0$), is detailed in~\cite{2015PhRvD..91h3535G}. An abbreviated description follows. The test statistic is the sum of two independent quantities $T_s$ and $T_b$: the sum of the weights of events due to dark matter (signal) and all other sources (background). The weights of individual signal events are statistically independent and they are independent of the weights of background events. Further, in this study we assume that background events are all independent of each other.

Under these conditions, the variables $T_s$ and $T_b$ are described by compound Poisson distributions: the sum of independent random variables (the weights) where the number of terms in the sum is a Poisson distributed variable. All that is required to construct the distribution is the expected number of events that will be detected with each weight. This is found by discretizing the $(\nu, E, \theta)$ space in a finite number of bins and computing the expected number of events in each bin (Eq.~\ref{eqn:s}) and the weight assigned to events in each bin (Eq.~\ref{eqn:wdef}).  Then a histogram is formed over the weight variable.

\begin{table}[t]
	\begin{center}
	\begin{tabular}{ l | c | c | c | c }
	Dwarf &  Zenith   & Azimuth &  Exposure & Energy Range   \\
	 & [deg] & [deg] & [hours]  &  [GeV] \\
	\hline
	Segue 1    	&  15-35 	& 100-260 	&  92.0 &  80 - 50000\\
	Draco      	&  25-40 	& 320-40		&  49.8  & 120 - 70000 \\
	Ursa Minor &  35-45 	& 340-30 		&  60.4 &  160 - 93000\\
	Bo{\"o}tes 1     	&  15-30 	& 120-249 	&  14.0  & 100 - 41000\\
	Willman 1  &  20-30 	& 340-40 		&  13.6  & 100 - 43000\\
	\end{tabular}
	\caption{Dwarf galaxy zenith and azimuth range, total accumulated exposure and energy range after cuts are applied. Azimuth is measured east from north. Upper energy range is defined as energies where the uncertainty in effective area is less than 10\%. }
	\label{tab:dwarfs}
	\end{center}
\end{table}

For the background events we consider the same discretized $(\nu, E, \theta)$ space. The weight of events in each bin is computed as above. The expected number of background events in each bin is computed using the empirical energy distribution of the OFF events and assuming the background events will be isotropic within the ON region. Specifically, each OFF event from dwarf $\nu$ with reconstructed energy in bin $E$ contributes $\alpha d\Omega_j / \Omega$ expected events to the $(\nu, E, \theta_j)$ bin, where $\alpha$ is the ON/OFF ratio for the run, $d\Omega_j$ is the solid angle of the $j$-th $\theta$-bin, and $\Omega$ is the total solid angle of the ON region.  This procedure is equivalent to a background model where events are sampled from OFF regions (with replacement) and distributed isotropically within the ON region; the probability of selecting an OFF event is proportional to its $\alpha$ value. 

The probability distribution for $T$ is the convolution of the probability distributions for $T_s$ and $T_b$ (since $T = T_s + T_b$). The compound Poisson distributions and the convolutions are efficiently calculated using standard Fast Fourier Transform techniques. 

In principle, the statistical power of the analysis can be increased by having an event\textsc{\char13}s weight depend on the run in which it was detected (in addition to it’s energy, angular separation, and which dwarf field it was detected in). This generalization would automatically and optimally ``downgrade'' runs which had poor observing conditions (smaller effective area, larger background flux). However, this requires having accurate background models and response functions on a run by run basis and current datasets are not large enough to allow this. In general, the search becomes more sensitive as the event weights are allowed to depend on more observables.

\section{Results}

\subsection{Search for annihilation in individual dwarfs}
The search for dark matter annihilation is performed by measuring $\Tobs$ and comparing this with the probability distribution for $T$ due to background. A search in an individual dwarf field is performed by setting the weights of events from all other dwarfs to zero. The weight function Eq.~\eqref{eqn:wdef} requires a signal hypothesis $s(\nu, E, \theta)$ which depends on the dark matter parameters $M$, $\sigv$, and $B_i$. 
We perform a search for dark matter of each mass and annihilation channel (assuming $B_i = 1$) in heavy quarks ($b\bar{b}$) and leptons ($\tau^{+}\tau^{-}$) as well as a two photon final state. The cross section $\sigv$ is a measure of the expected signal amplitude and must be specified in order to assign weights. A specific value $\sigv_{90}$ is used: it is the value of the cross section for which there is a 90\% chance of making a 3$\sigma$ detection, where $\sigma$ is defined as number of standard deviations above the background. In VHE astronomy, 5$\sigma$ is typically required for a discovery. In practice, the search is essentially independent of the specific value of $\sigv$ used in the weighting, but $\sigv_{90}$ is chosen to make the search as sensitive as possible to cross sections that are on the verge of being detectable by the instrument.

\begin{figure}
	\centering
	\includegraphics[scale=0.9]{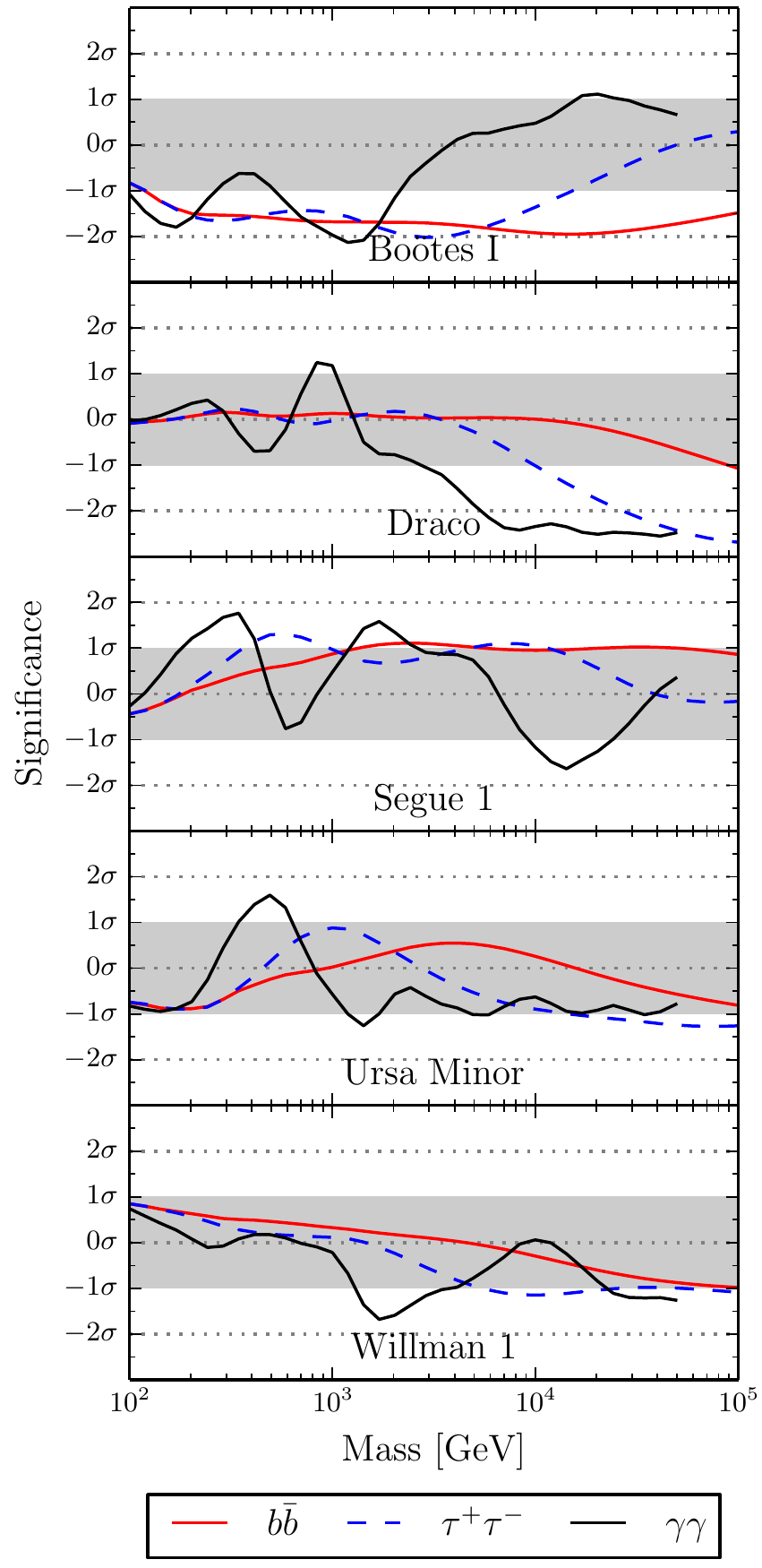}
	\caption{Results of the individual search for dark matter annihilation for three Standard Model final states. For each dark matter mass ($x$-axis), the $y$-axis gives the significance of detection, defined as the quantile of the probability distribution of the background-only model. This probability is converted into a ``sigma value'' using the inverse CDF of a standard Gaussian. The gray band represents the range of $\pm$1$\sigma$. }
	\label{fig:search_indiv}
\end{figure}

Figure~\ref{fig:search_indiv} shows the results for the search in the individual dwarfs. No evidence of dark matter annihilation at any mass has been observed in any one of the dwarfs. Note that annihilation into a two photon final state terminates at the highest energy of the event sample as shown in the last column of Table~\ref{tab:dwarfs}. These run from the lowest reconstructed energy for an off source event to an upper energy where the uncertainty in the effective area is 10\%. The limits given here are insensitive to these energy thresholds.

\subsection{Flux upper limits}

Due to the lack of any detectable signal and in order to compare with complementary experiments we derive a flux upper limit $\Phi_\gamma(E > \Emin)$, as 
\begin{eqnarray} 
\Phi_\gamma(> \Emin) &=& N_{\gamma, {\mathrm{obs}}} (>\Emin) \int_{\Emin}^\infty \frac{dN_\gamma}{dE} \, dE  \nonumber \\ 
&\times & \left[\sum_j \int_{\Emin}^\infty  \tau_j {\Aeff}_{,j}(E)\, \frac{dN_\gamma}{dE} \,dE \right]^{-1} 
\end{eqnarray}
where $N_{\gamma, {\mathrm{obs}}}$ is the total observed number of events along the direction of a dwarf, $\tau_j$ and $ {\Aeff}_{,j}(E)$ are the observation time and effective area of each $j$ run, respectively, and  $dN_{\gamma} / dE$ is the assumed source differential energy spectrum. 
The energy threshold $\Emin$ is defined here as the maximum of the efficiency curve which is defined as the effective area curve multiplied by the assumed source differential spectrum. In this case, the assumed differential spectrum is a power law of index -2.4.
The bounded profile likelihood ratio statistical method of Rolke et al. \cite{2005NIMPA.551..493R} is used in this  analysis to determine the upper limit on the number of gamma rays from the direction of each dwarf. The last column in Table~\ref{tab:dwarfsresult} shows the resulting upper limits.


\begin{table*}[t]
	\begin{center}
	\begin{tabular}{ l | c | c | c | c | c | c | c | c}
	Dwarf & $N_{ON}$ & $N_{OFF}$ & $\bar{\alpha}$ & Significance & $N^{95\%}$ & $\Phi^{95\%}$                          & $D$   & $\log_{10} J(0.17^{\circ})$ \\
	          & [counts]   & [counts]     &                        & [$\sigma$]  & [counts]        & [10$^{-12}$cm$^{2}$s$^{-1}$] & [kpc] & [GeV$^{2}$ cm$^{-5}$] \\
	\hline
	Segue 1   	 & 	15895  &	120826	&	0.131	&  0.7   	& 235.8	& 0.34  & 23  & 19.2$^{+0.3}_{-0.3}$ \\
	Draco      	 &   	 4297   &	 39472	&  	0.111	& -1.0	& 33.5	& 0.15  & 76  & 18.3$^{+0.1}_{-0.1}$ \\
	Ursa Minor &      4181   &	 35790	&	0.119	& -0.1   	& 91.6	& 0.37  & 76  & 18.9$^{+0.3}_{-0.3}$ \\
	Bo{\"o}tes 1 &  	 1206   &	 10836	&	0.116	& -1.0 	& 34.5	& 0.40  & 66  & 18.3$^{+0.3}_{-0.4}$ \\
	Willman 1   &     1926   &	 18187	&	0.108	& -0.6    	& 23.5	& 0.39  & 38  & N/A\\
	\end{tabular}
	\caption{Dwarf galaxy detection significance (generalized Li \& Ma method) and integral flux upper limit with 95\% confidence level above 300 GeV, assuming a spectral index of -2.4. The last two columns are the heliocentric distance to each object and the inferred value of $J$-profile integrated within a cone with half-angle of $0.17^{\circ}$ (i.e. over the ON region), errors denote the 16th and 84th percentiles on the posterior \cite{2015ApJ...801...74G}. Note that this analysis uses the $J$-profile convolved with the VERITAS instrument response as discussed in Section IV.} 
	\label{tab:dwarfsresult}
	\end{center}
\end{table*}

\subsection{Combined search}
Compared with examining individual dwarfs, pooling the data from all of them yields a search sensitive to weaker annihilation cross sections. The ON events from Bo\"{o}tes 1, Draco, Segue 1, and Ursa Minor are weighted according to Eq.~\eqref{eqn:wdef} and summed according to Eq.~\eqref{eqn:Tdef}. We do not include Willman 1 in the joint analyses because its irregular kinematics preclude a reliable determination of its $J$-profile via the Jeans equation (see discussion in Section III and \cite{2015ApJ...801...74G}).

In this approach, the $J$-profiles must be taken into account since they are no longer degenerate with the cross section. We incorporate the systematic uncertainties in the dark matter distributions in the dwarfs by performing an ensemble of searches. For each, we assign each dwarf a $J$-profile from the posterior distribution of halo parameters~\cite{2015PhRvD..91h3535G}. The scatter of the search resulting from many such realizations gives a measure of the systematic uncertainty due to our incomplete understanding of the density profiles in the dwarfs.

\begin{figure*}
	\centering
	\includegraphics[scale=0.68]{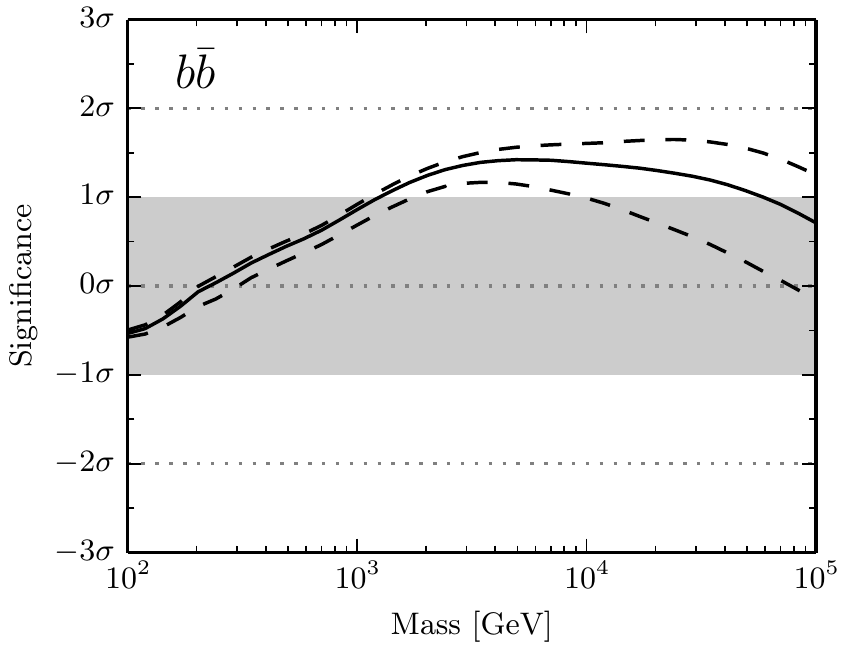}
	\includegraphics[scale=0.68]{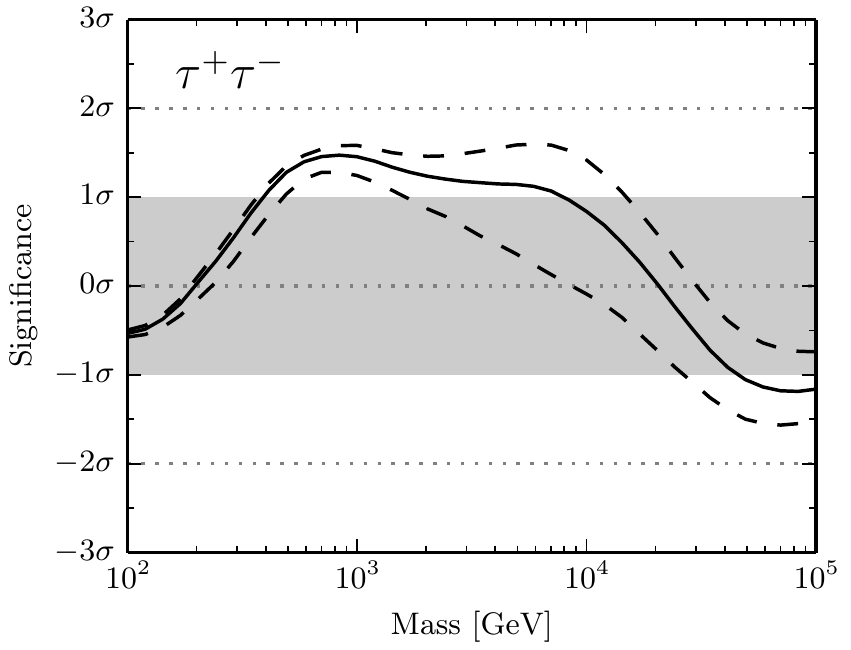}\\
	\includegraphics[scale=0.68]{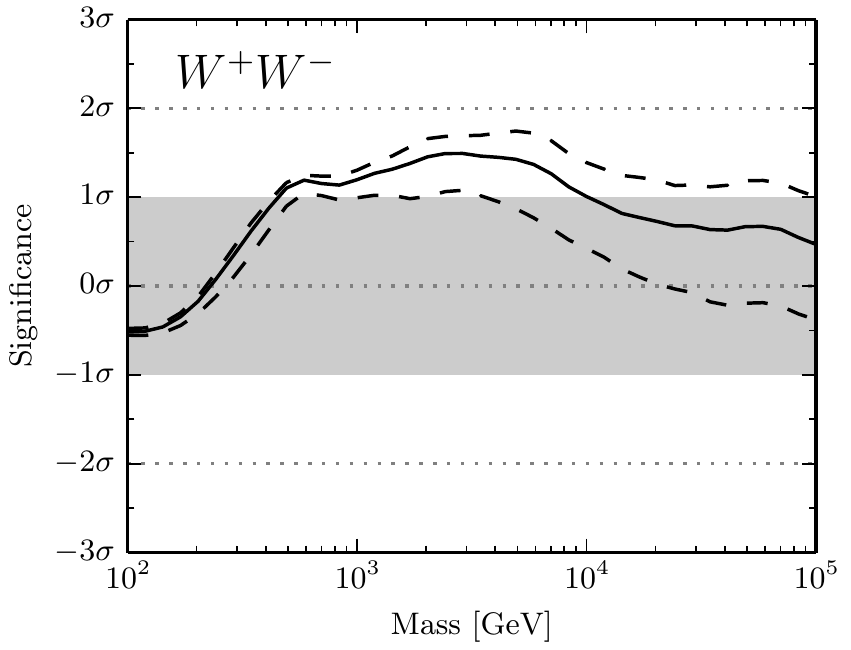}
	\includegraphics[scale=0.68]{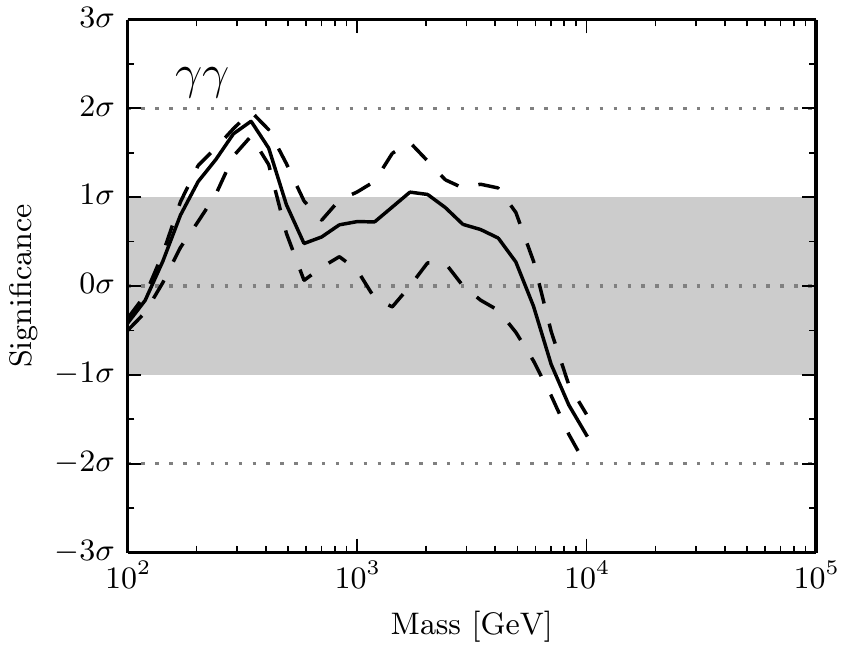}\\
	\caption{Results of the combined search for dark matter annihilation in the four dwarf galaxies whose dark matter density profiles can be reliably determined for annilhilation into four standard model final states. For each dark matter mass ($x$-axis), the $y$-axis gives the significance of detection, defined as the quantile of the probability distribution of the background-only model. This probability is converted into a ``sigma value'' using the inverse CDF of a standard Gaussian. The dashed lines show how the detection significance depends on the uncertainty in the dark matter density profiles (the solid line is the median over all allowed density profiles).}
	\label{fig:search_combined}

\end{figure*}

The results of the combined search are shown in Figure~\ref{fig:search_combined}. The dashed lines bound 68\% of the halo profile realizations and the solid line is the median significance. The combined observation shows no sign of dark matter annihilation in any channel.

\subsection{Upper limits on the cross section}

\begin{figure*}
\centering
\includegraphics[scale=0.68]{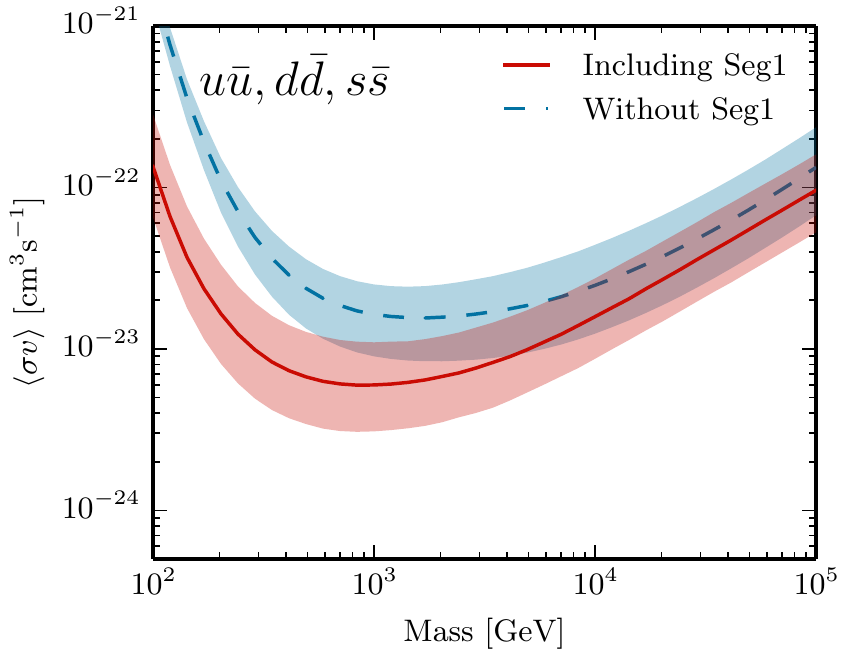}
\includegraphics[scale=0.68]{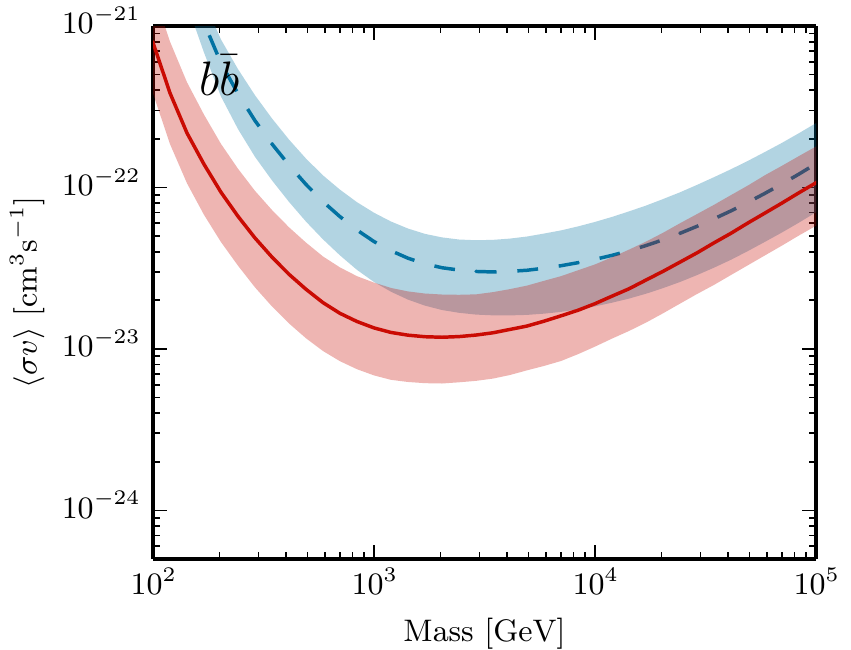}
\includegraphics[scale=0.68]{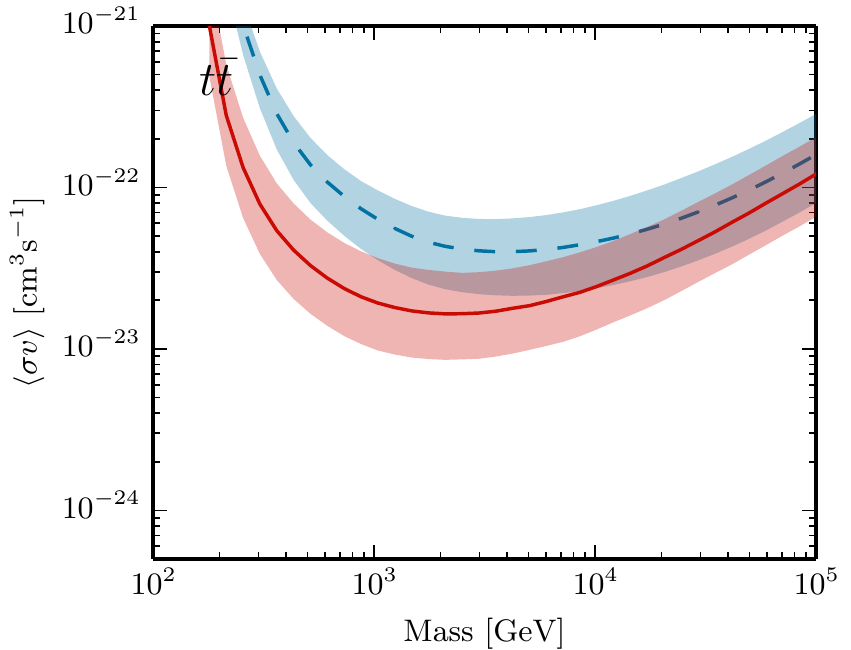}\\
\includegraphics[scale=0.68]{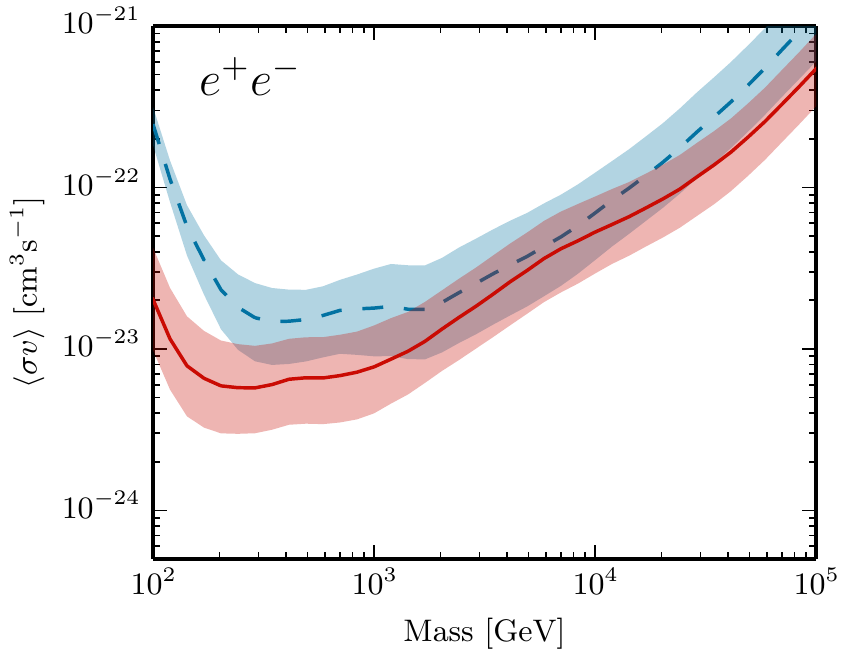}
\includegraphics[scale=0.68]{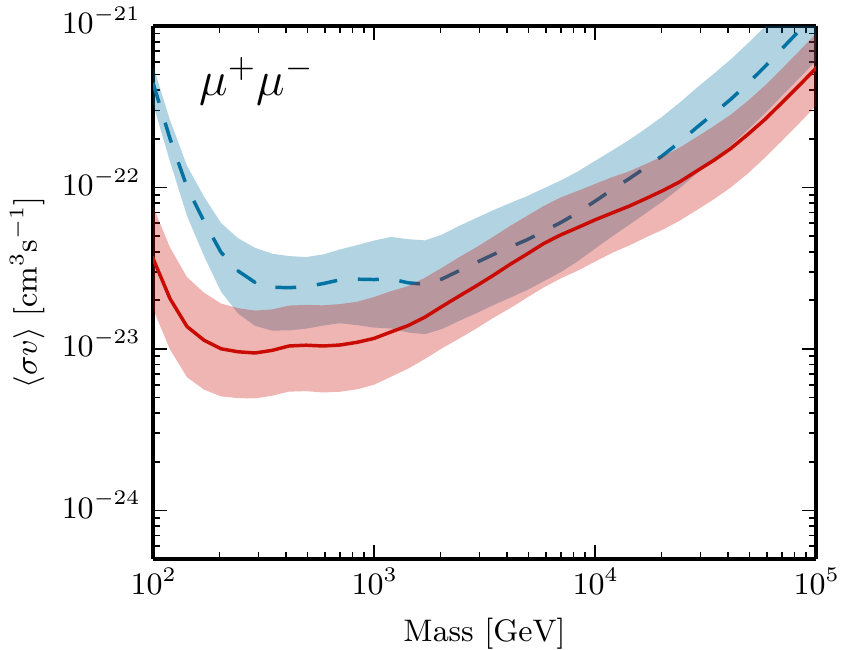}
\includegraphics[scale=0.68]{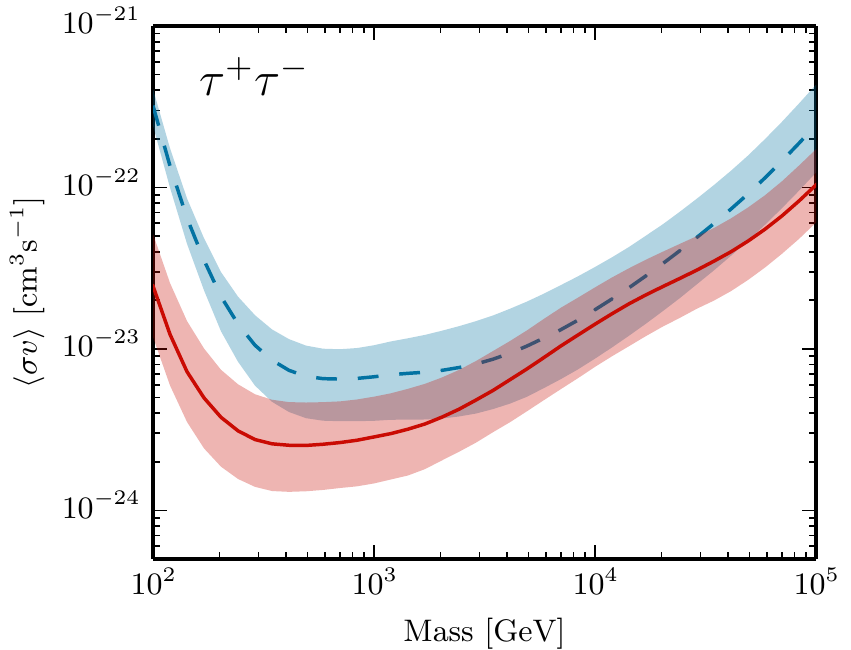} \\
\includegraphics[scale=0.68]{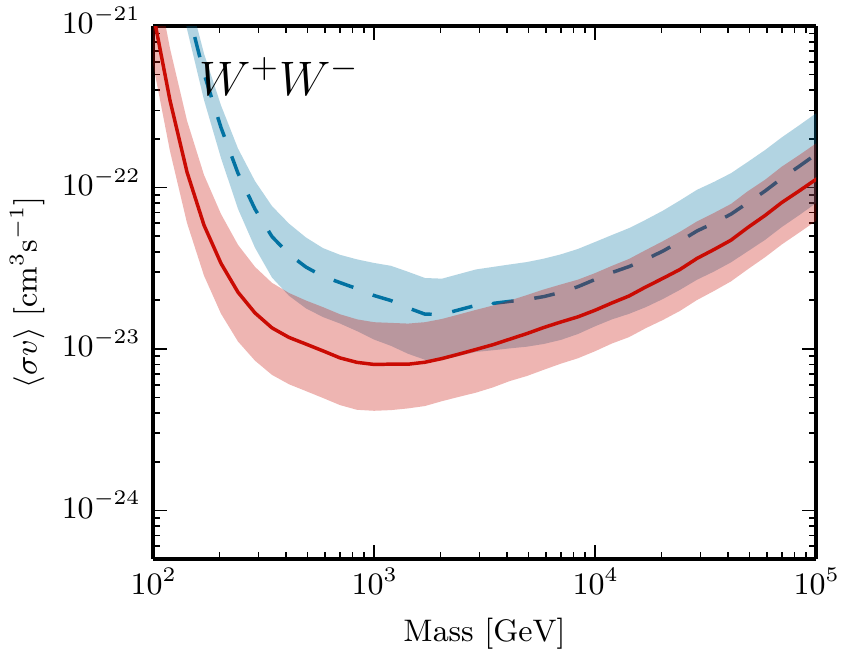}
\includegraphics[scale=0.68]{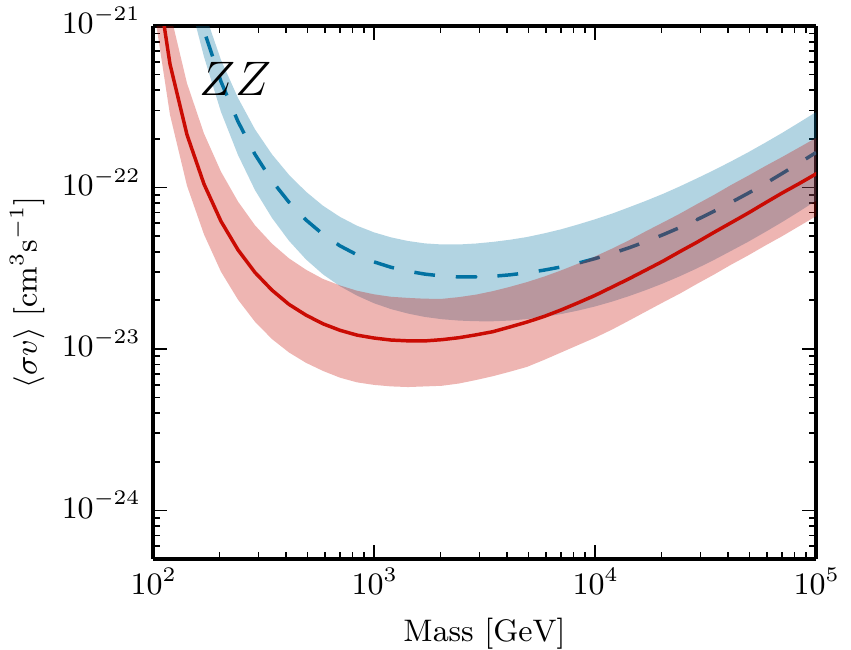}
\includegraphics[scale=0.68]{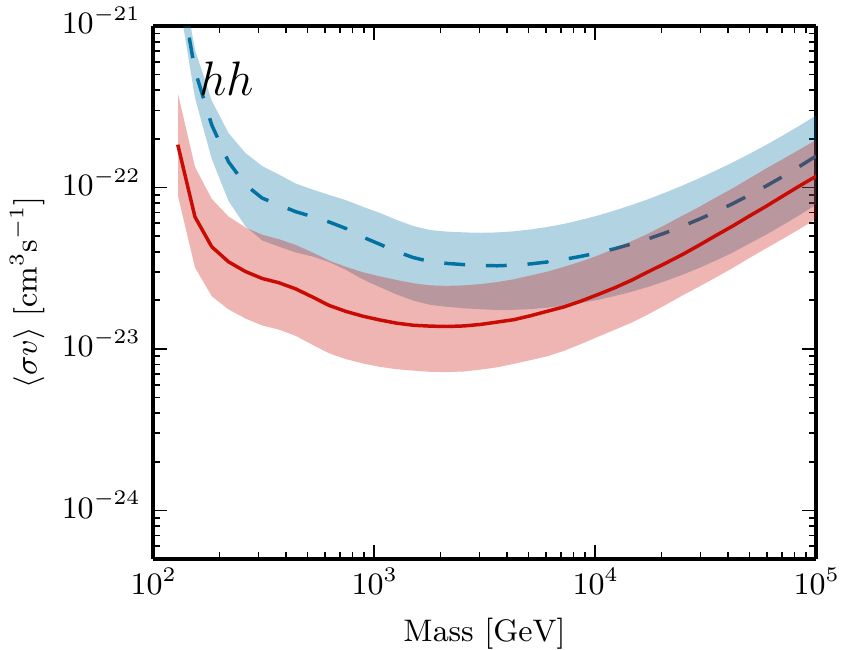} \\ 
\caption{Annihilation cross section limits from the joint analysis of dwarf galaxies.  The shaded bands are the systematic $1 \sigma$ uncertainty in the limit derived from many realizations of halo $J$-profiles of the dwarfs consistent with kinematic data. The solid line depicts the median of this distribution of limits over the halo realizations with all dSphs except Willman~1. The dashed line depicts the median limits of the distribution of limits without Segue~1 and Willman~1. A machine-readable file tabulating these limits is available as supplemental material. }
\label{fig:combinedlimits_syst}
\end{figure*}

\begin{figure*}
\centering
\includegraphics[scale=0.68]{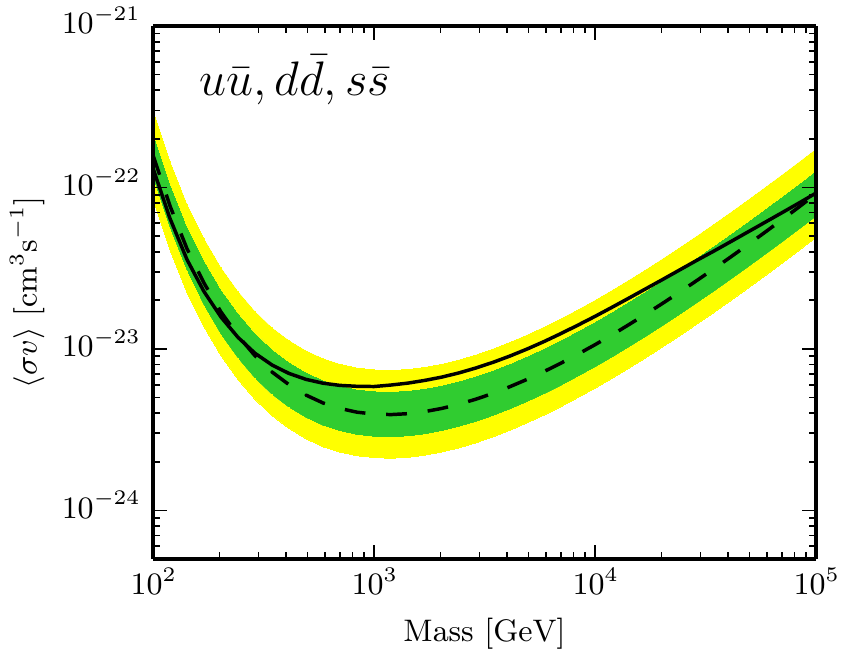}
\includegraphics[scale=0.68]{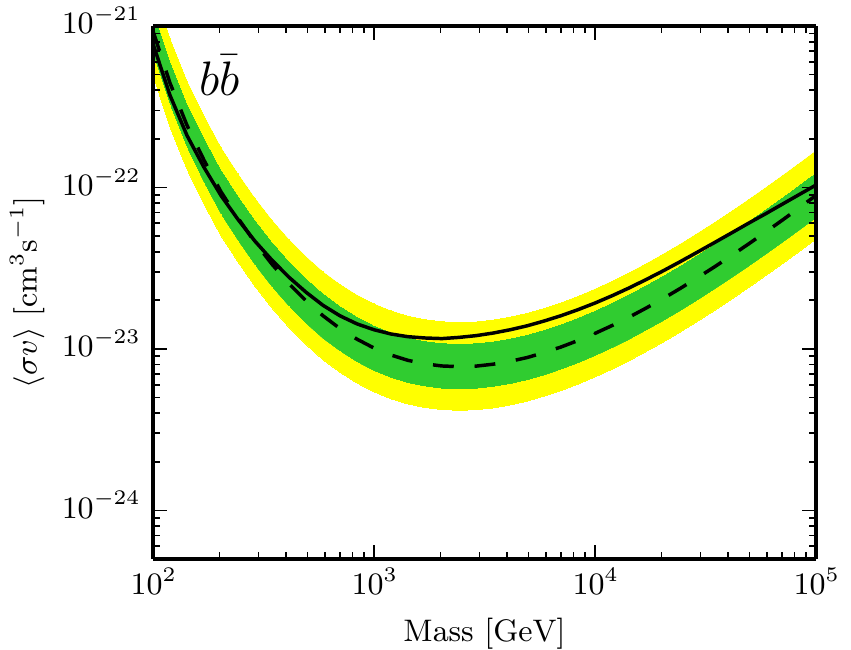}
\includegraphics[scale=0.68]{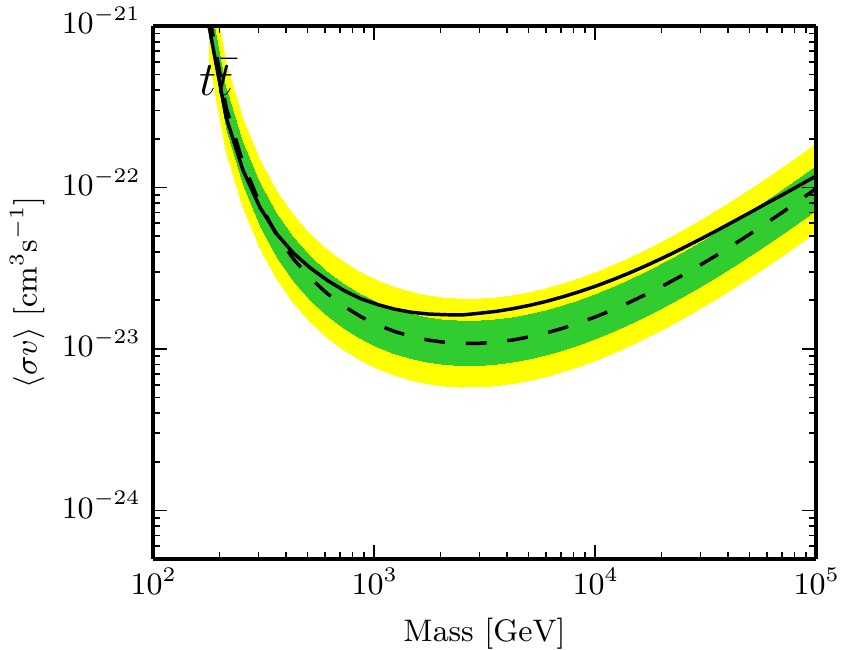}\\
\includegraphics[scale=0.68]{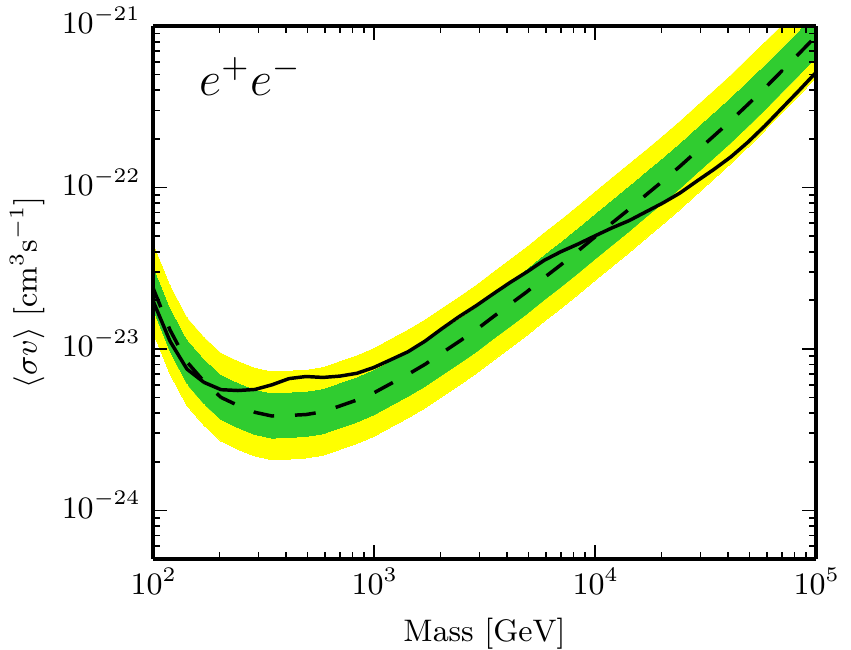}
\includegraphics[scale=0.68]{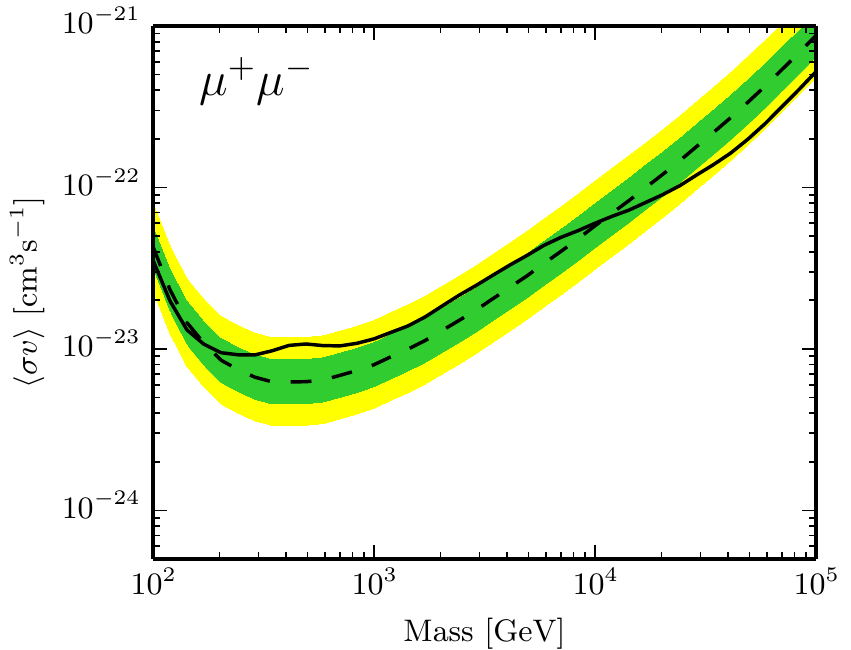}
\includegraphics[scale=0.68]{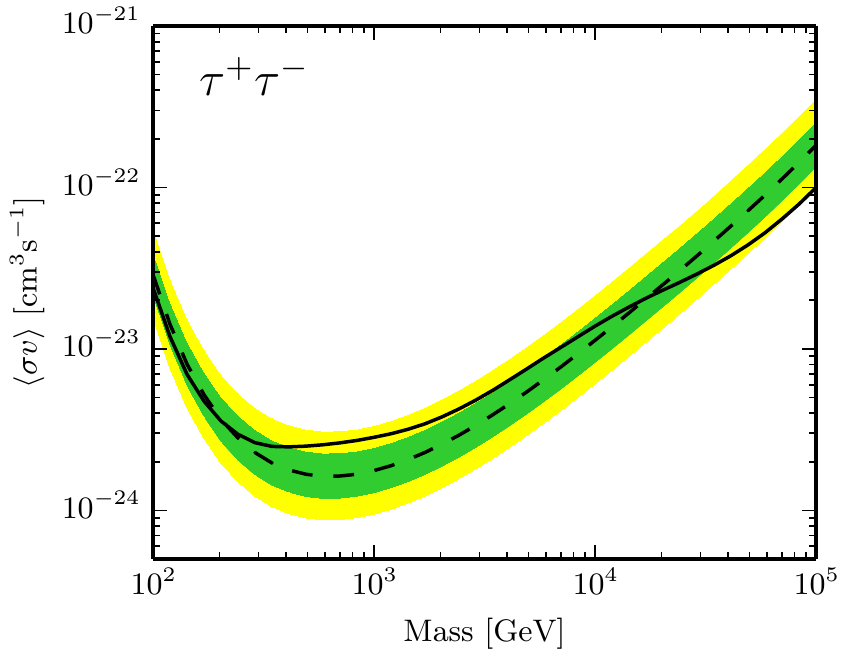} \\
\includegraphics[scale=0.68]{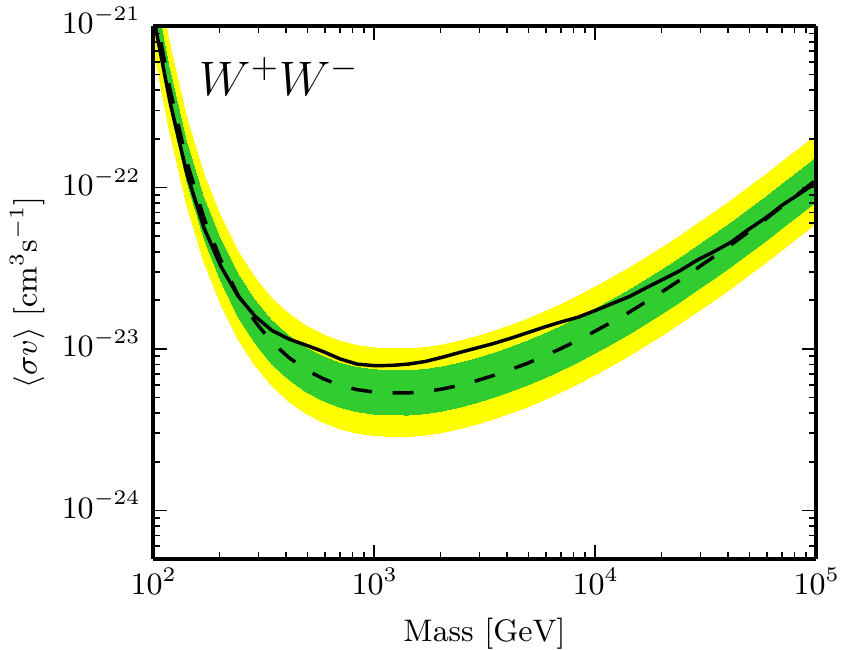}
\includegraphics[scale=0.68]{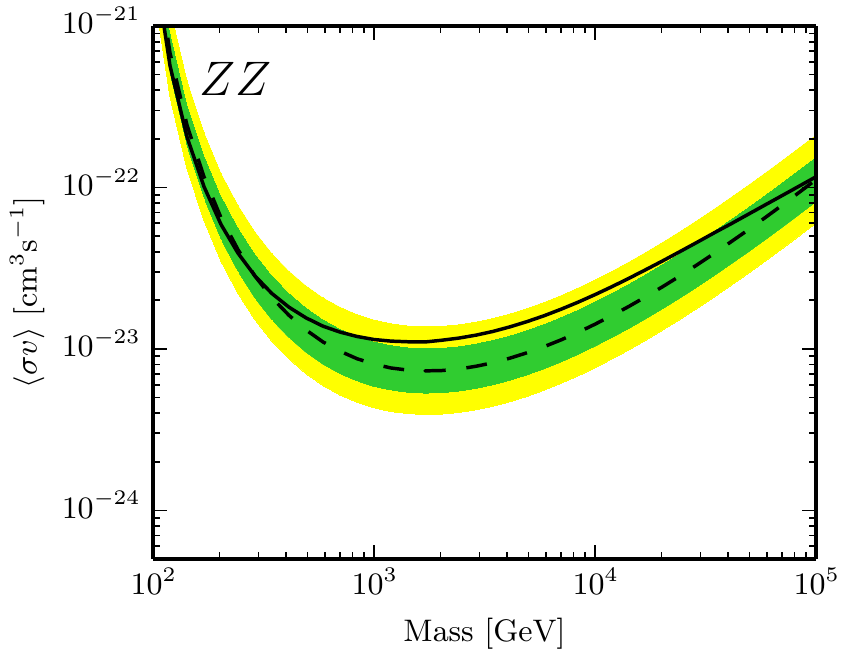}
\includegraphics[scale=0.68]{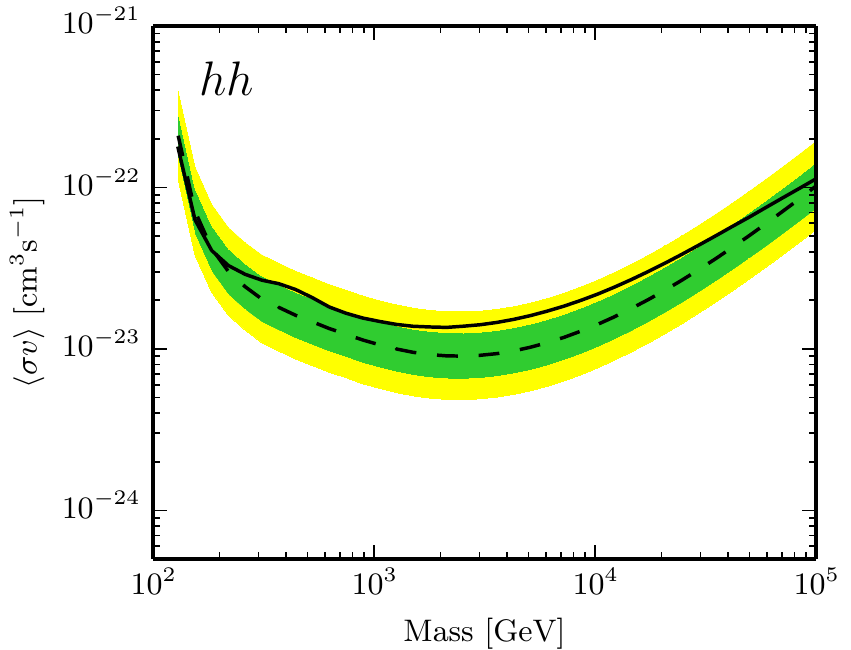} \\ 
\caption{Expected annihilation cross section limits from the joint analysis of four dwarf galaxies.  The green and yellow bands depict the a 68\% and 95\% chance of the limit being in these regions. The expected limit has a 50\% chance to be below the dashed line, while the solid line shows the observed upper limit for a particular realization of halo density profile (the actual width spanned by the complete sample of different profiles is shown as the shaded area in each panel of Figure \ref{fig:combinedlimits_syst}).
\label{fig:combinedlimits_exp}}
\end{figure*}

\begin{figure*}
	\centering
	\includegraphics[scale=0.68]{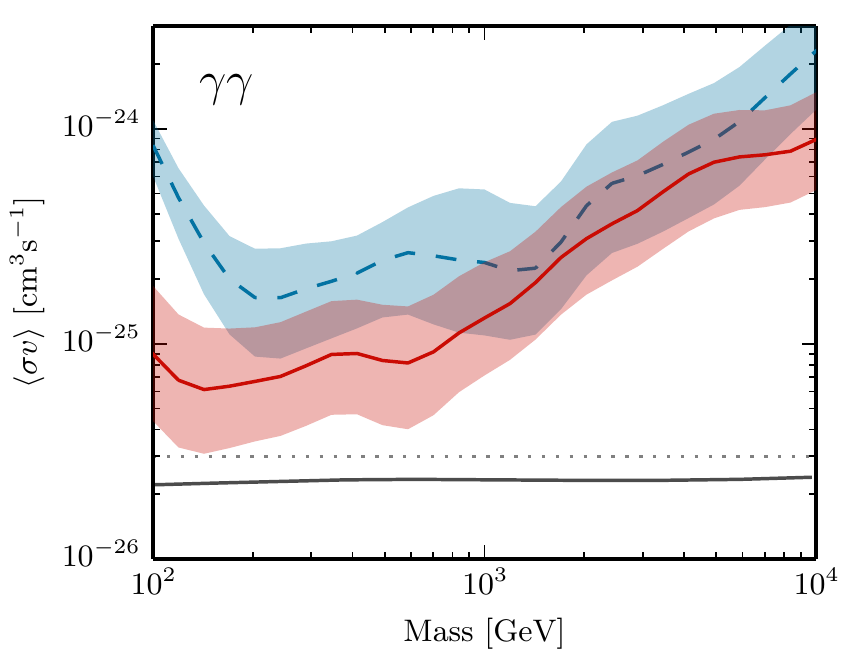}
	\includegraphics[scale=0.68]{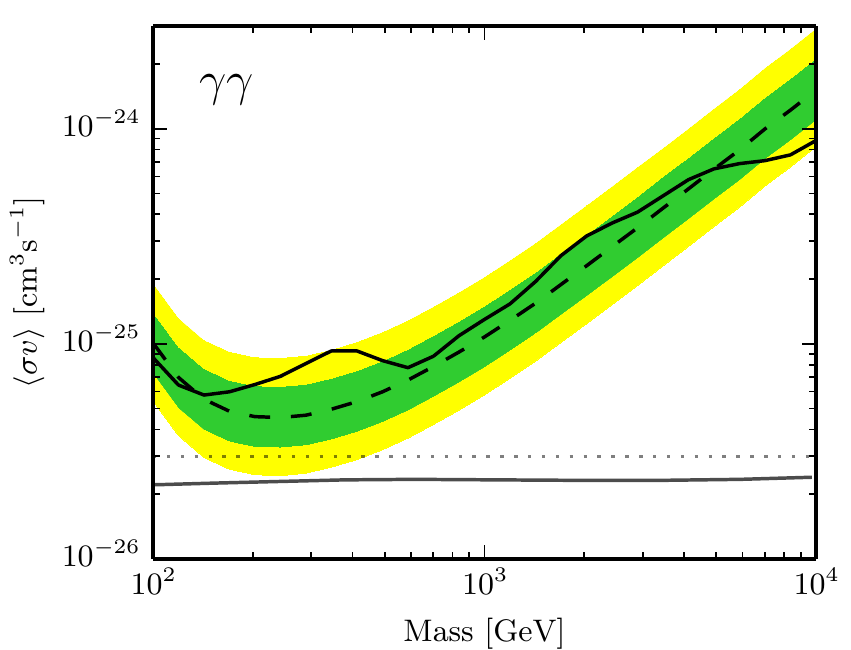}
	\caption{Same as Figures~\ref{fig:combinedlimits_syst} \&~\ref{fig:combinedlimits_exp} for the case of dark matter annihilation to a two photon final state. }
	\label{fig:line}

\end{figure*}

We slightly modify the procedure of~\cite{2015PhRvD..91h3535G} to compute cross section upper limits. In that work 95\% confidence limits were generated using the Neyman construction of confidence belts. There, a hypothesis test is performed at every value of the cross section. The $\sigv$-space is divided into two regions where the hypothesis can and cannot be rejected at 95\% confidence, with high enough values of $\sigv$ always being rejected. The boundary between the regions constitutes a 95\% upper limit on the cross section. The hypothesis test is performed by asking, for a given value of $\sigv$, whether the probability that $T < \Tobs$ is less than 5\%. If it is, then this value of the cross section is rejected.

In this work we adopt the $\CLs$ technique~\cite{1999NIMPA.434..435J,2002JPhG...28.2693R} (sometimes called modified frequentist analysis) to produce upper limits. This method is strictly more conservative than the Neyman construction described above, i.e. always gives a larger upper limit, but has the benefit of being immune to downward fluctuations of background causing the upper limits to be much lower than the experimental sensitivity. That is, in the scheme described above, if there is a strong enough negative fluctuation of background so that $\prob(T < \Tobs \vert \sigv=0) < 5\%$ even the $\sigv=0$ hypothesis will be rejected causing the $\sigv$ upper limit to be zero.


The 95\% confidence level upper limits on the annihilation cross section are presented in Figures ~\ref{fig:combinedlimits_syst} and ~\ref{fig:line}. Each panel constrains dark matter with a 100\% branching fraction into various Standard Model final states. The shaded band represents the $1\sigma$ systematic uncertainty induced by our imperfect knowledge of the dwarfs' density profiles. They are produced by repeating the limit calculation over an ensemble of realizations of the dwarf halos from the distribution described in Section~\ref{sec:Jprofiles}. The lower, upper, and center of the band correspond to the 16th, 84th, and 50th percentiles of the distribution of limits over halo realizations. All other systematic uncertainties are negligible in this work in comparison and have been ignored. 

\begin{figure}
	\centering
	\includegraphics[scale=0.9]{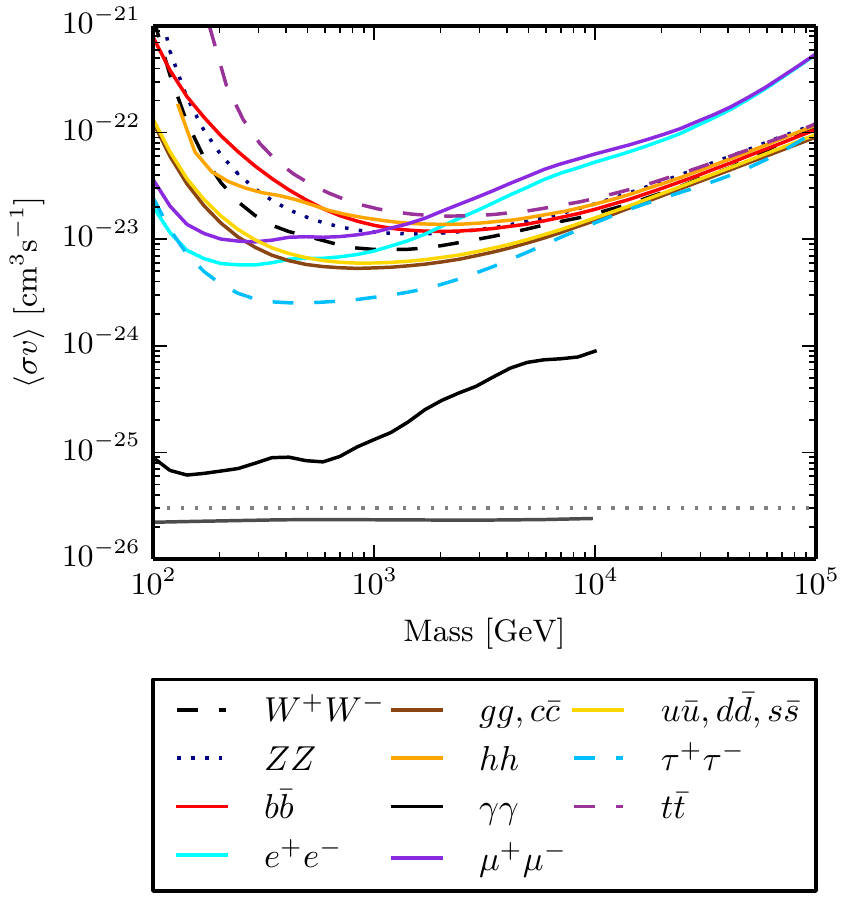}
	\caption{The median annihilation cross section limit from all dwarf galaxies and for all channels (the solid curves of Figure~\ref{fig:combinedlimits_syst} and ~\ref{fig:line}).  The strongest continuum constraints are from a heavy lepton final state.  The thin dashed horizontal line corresponds to the benchmark value of the required relic abundance cross section ($3 \times 10^{-26} {\mathrm{cm}}^3/{\mathrm{s}}$), while the solid horizontal line corresponds to the detailed calculation of this quantity  \cite{2012PhRvD..86b3506S}. }
	\label{fig:limits_all_channels}

\end{figure}

As discussed in Section III, recent work has questioned the reliability of the $J$-profile of Segue~1 because of possible foreground contamination of its spectroscopic sample. By excluding Segue~1 from the combined analysis (i.e. setting its dark matter density to zero) we can bracket the effect that this unmodeled systematic uncertainty has on the particle physics constraints. Cross section limits are substantially weakened below a particle mass of about 400~GeV due to the lower energy threshold for the Segue~1 observations as compared to Draco and Ursa Minor (see Figure~2). Depending on the annihilation channel, excluding Segue~1 increases the $\sigv$ limit by a factor between 9-14 at 100 GeV, 4-7 at 200 GeV, 2-5 at 400 GeV, 2-3.3 at 1~TeV, and 1.2-2 above 10~TeV. Combined limits with and without Segue 1 included in Figures ~\ref{fig:combinedlimits_syst} and ~\ref{fig:line}.

\subsection{Statistical fluctuations}

Hypothetically, if we were to repeat the measurement many times while holding the $J$-profiles of the dwarfs fixed, we would still obtain a distribution of limits due to statistical fluctuations intrinsic to a finite data set. We quantify the impact of the statistical uncertainty by looking at the distribution of the test statistic under the background-only hypothesis. That is, without using the events in the ON region, we take $\Tobs$ to be a given quantile of $\prob(T \mid \sigv=0)$ and find the upper limit that would be obtained if this value had actually been measured. By taking the  $0, \pm 1\sigma, \pm 2\sigma$ quantiles we find ranges where the observed limit is likely to lie. These are plotted in Figures ~\ref{fig:combinedlimits_exp} and ~\ref{fig:line}. Specifically, due to random fluctuations of the background in the ON region, there is a 68\% chance that the observed limit lies in the green band and a 95\% chance that it lies in the yellow band. The dashed line is the median expected limit: there is a 50\% chance that the observed limit is stronger than this. The solid black curve is the observed limit using the data from the ON region. This plot contains similar information to Figures ~\ref{fig:search_indiv} and~\ref{fig:search_combined}. It shows how consistent the observations are with the background-only hypothesis. These plots were made using a particular set of $J$-profiles for the dwarfs, chosen to align well with Figures~\ref{fig:combinedlimits_syst} and ~\ref{fig:line}, and are meant to illustrate the experimental sensitivity of VERITAS and show the effect of background fluctuations on the cross section limits. The median limits for all channels are shown in Figure ~\ref{fig:limits_all_channels}.

\begin{figure*}
	\centering
	\includegraphics[scale=0.4]{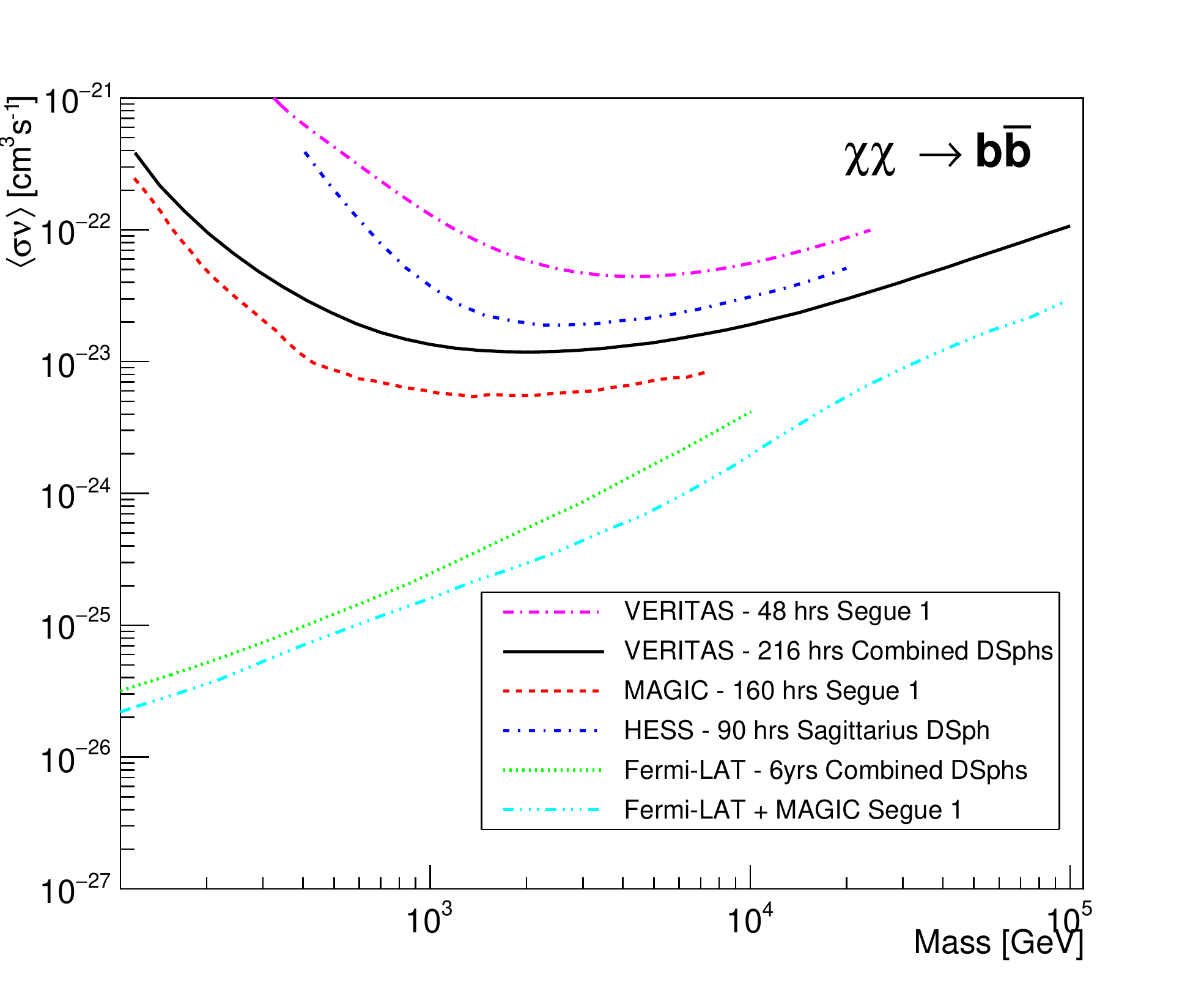}
	\includegraphics[scale=0.4]{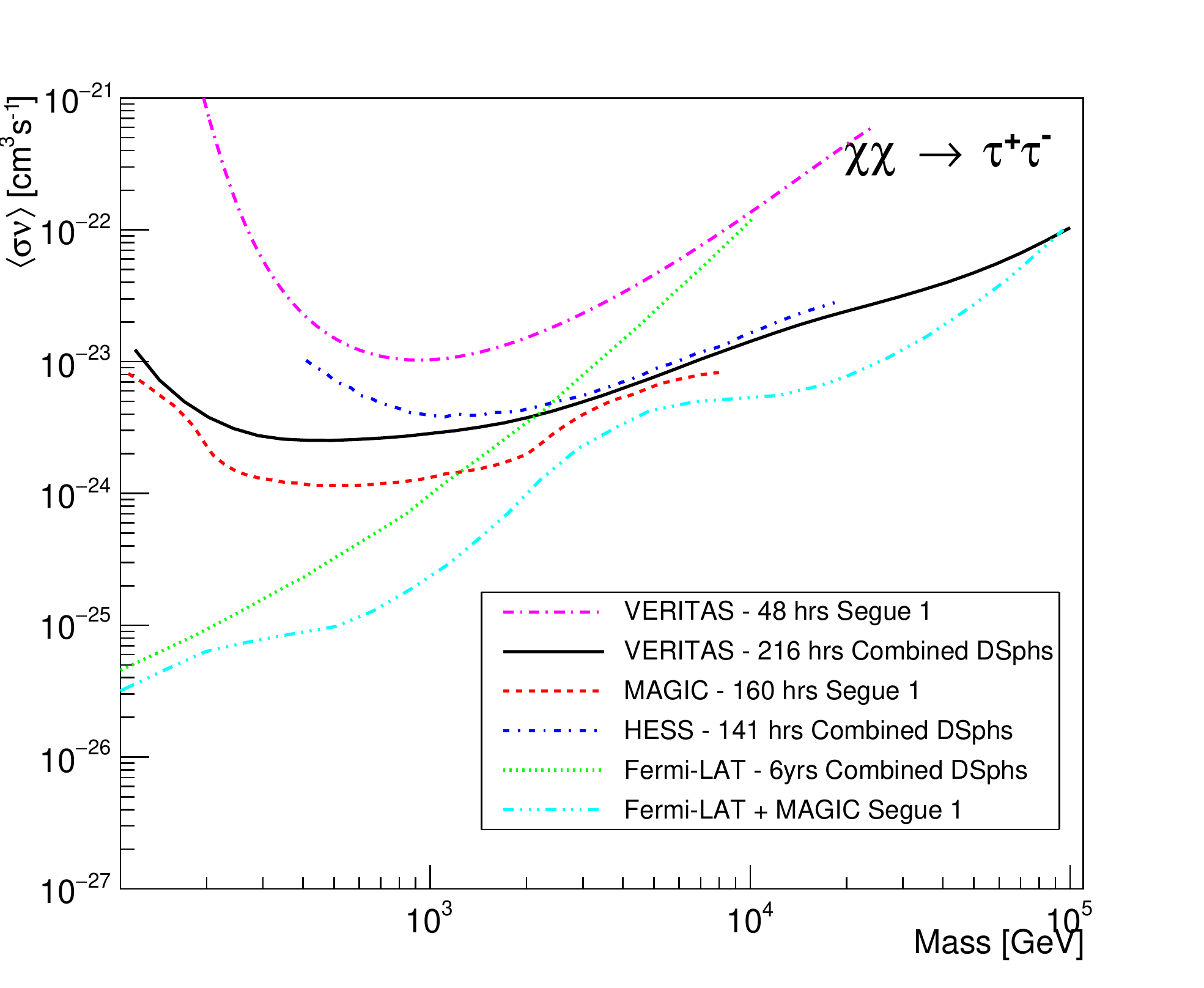}
	\caption{Annihilation cross section limits for dwarf spherioidal galaxies from this work, HESS \cite{2014PhRvD..90k2012A}, MAGIC \cite{2014JCAP...02..008A}, Fermi-LAT \cite{2015PhRvL.115w1301A}, a combined result of MAGIC and Fermi-LAT \cite{1475-7516-2016-02-039} as well as previous VERITAS results  \cite{PhysRevD.91.129903} for the $b\bar{b}$ (left) and $\tau^{+}\tau^{-}$ (right) channels. }
	\label{fig:gamma_ray_limits}

\end{figure*}

\section{Conclusions and Discussion}

The VERITAS limits in comparison with other concurrent gamma-ray instruments as well as older VERITAS results are shown in Figure~\ref{fig:gamma_ray_limits}. For the first time in an IACT DM search, this work uses the individual direction in addition to energy information of each event in the construction of the test statistic. The VERITAS results shown in this work are a substantial improvement over the entire WIMP mass range over the previous result with 48 hours on Segue 1 \cite{PhysRevD.91.129903}. VERITAS has a diverse dark matter program: observing time is divided between both the classical and ultrafain dSphs since we still have an imperfect knowledge of dwarf spheroidals and their J-profiles and their systematic uncertainties. This is especially important in light of the considerable uncertainty in the reconstruction of dwarf dark matter density profiles (see Section III and Figure 5). The strategy taken here of combining multiple targets in a single dark matter search mitigates sensitivity to future findings about particular galaxies. Pointed telescopes that rely heavily on a single target (e.g. Segue 1) may find their results susceptible to large, unaccounted systematic uncertainties. The Fermi-LAT, with a large duty cycle on all dSphs and low backgrounds, sets more stringent limits in the low mass range; however, the IACTs (VERITAS, MAGIC and HESS) put more stringent limits at the high mass range ($M\gtrsim$ 1 TeV), where Fermi-LAT has very low statistics. 

Although no future hardware upgrades are currently planned for VERITAS, several advanced analysis techniques are starting to be deployed for VERITAS data. These techniques (e.g. boosted decision trees for $\gamma$/Hadron separation\cite{2017APh....89....1K}) could boost dark matter sensitivity by 30-50\%. Additionally, the cuts used for this analysis were ``point-like'', optimized for the detection of point sources. Nearly all the dark matter profiles for dwarf galaxies extend larger than the ON source region used in this work. An extended source analysis using a larger signal region could boost dark matter sensitivity by as much as a factor of two, dependent on the J-profile for each dSph. Dwarfs and other dark matter targets remain high-priority targets for the remainder of the lifetime of VERITAS.

The current upper limits on the annihilation cross section are about two orders of magnitude away from the relic abundance value ($\sigv \approx 10^{-26} \cm^2 \second^{-1}$). This highlights the importance of improving both the instrumental sensitivity and the particle physics analysis. It is vital to extract all information present in the data to push experiments to the limit of their capability. The event weighting method, applied to IACT analysis for the first time, is a powerful and efficient way to combine multiple data sets and use our knowledge of the dark matter distribution and particle properties to perform optimal searches. For the first time, the event angular direction is used in addition to the energy of individual events for an IACT dark matter search.

It should be noted that the dark matter annilihation limits in this work were independently cross checked with a variation of the Full Likelihood utilized by the MAGIC collaboration\cite{2014JCAP...02..008A} for a single halo realization for each dSph. The only major difference is that DM profiles were convolved with the VERITAS PSF described in this work, giving an integrated $J$-factor that is a function of energy. The combined dwarf limits of the two methods agreed within both the expected limits and $J$-factor systematic limits for the entire DM mass range used in this work. 

To reach the thermal relic cross section, it may be necessary to combine all data taken from several gamma-ray telescopes into a single, deep search, expanding on the example that has been demonstrated by the MAGIC and Fermi-LAT collaborations \cite{1475-7516-2016-02-039}. The methods we employed here may help prepare the experimental astroparticle physics community to accomplish this with upcoming experiments such as the Cherenkov Telescope Array (CTA) \cite{2013APh....43....3A}.

\section{Acknowledgments}
This research is supported by grants from the U.S. Department of Energy Office of Science, the U.S. National Science Foundation and the Smithsonian Institution, and by NSERC in Canada. We acknowledge the excellent work of the technical support staff at the Fred Lawrence Whipple Observatory and at the collaborating institutions in the construction and operation of the instrument. 
SMK acknowledges support from the Department of Energy through Grant DE-SC0010010, and thanks the Aspen Center for Physics and the Center for Experimental and Theoretical Underground Physics (CETUP$^*$) for hospitality where part of this work was completed. This research used computational resources of the National Energy Research Scientific Computing Center, a DOE Office of Science User Facility supported by the Office of Science of the U.S. Department of Energy under Contract No. DE-AC02-05CH11231. The VERITAS Collaboration is grateful to Trevor Weekes for his seminal contributions and leadership in the field of very high energy gamma-ray astrophysics, which made this study possible.

\bibliographystyle{apj}
\bibliography{manuscript}

\begin{thebibliography}{}
\expandafter\ifx\csname natexlab\endcsname\relax\def\natexlab#1{#1}\fi

\bibitem[{{Abramowski} {et~al.}(2014){Abramowski}, {Aharonian}, {Ait Benkhali},
  {Akhperjanian}, {Ang{\"u}ner}, {Backes}, {Balenderan}, {Balzer}, {Barnacka},
  {Becherini}, \& et~al.}]{2014PhRvD..90k2012A}
{Abramowski}, A., {Aharonian}, F., {Ait Benkhali}, F., {et~al.} 2014, \prd, 90,
  112012

\bibitem[{{Acharya} {et~al.}(2013){Acharya}, {Actis}, {Aghajani}, {Agnetta},
  {Aguilar}, {Aharonian}, {Ajello}, {Akhperjanian}, {Alcubierre},
  {Aleksi{\'c}}, \& et~al.}]{2013APh....43....3A}
{Acharya}, B.~S., {Actis}, M., {Aghajani}, T., {et~al.} 2013, Astroparticle
  Physics, 43, 3

\bibitem[{{Ackermann} {et~al.}(2015){Ackermann}, {Albert}, {Anderson},
  {Atwood}, {Baldini}, {Barbiellini}, {Bastieri}, {Bechtol}, {Bellazzini},
  {Bissaldi}, {Blandford}, {Bloom}, {Bonino}, {Bottacini}, {Brandt}, {Bregeon},
  {Bruel}, {Buehler}, {Caliandro}, {Cameron}, {Caputo}, {Caragiulo}, {Caraveo},
  {Cecchi}, {Charles}, {Chekhtman}, {Chiang}, {Chiaro}, {Ciprini}, {Claus},
  {Cohen-Tanugi}, {Conrad}, {Cuoco}, {Cutini}, {D'Ammando}, {de Angelis}, {de
  Palma}, {Desiante}, {Digel}, {Di Venere}, {Drell}, {Drlica-Wagner}, {Essig},
  {Favuzzi}, {Fegan}, {Ferrara}, {Focke}, {Franckowiak}, {Fukazawa}, {Funk},
  {Fusco}, {Gargano}, {Gasparrini}, {Giglietto}, {Giordano}, {Giroletti},
  {Glanzman}, {Godfrey}, {Gomez-Vargas}, {Grenier}, {Guiriec}, {Gustafsson},
  {Hays}, {Hewitt}, {Horan}, {Jogler}, {J{\'o}hannesson}, {Kuss}, {Larsson},
  {Latronico}, {Li}, {Li}, {Llena Garde}, {Longo}, {Loparco}, {Lubrano},
  {Malyshev}, {Mayer}, {Mazziotta}, {McEnery}, {Meyer}, {Michelson}, {Mizuno},
  {Moiseev}, {Monzani}, {Morselli}, {Murgia}, {Nuss}, {Ohsugi}, {Orienti},
  {Orlando}, {Ormes}, {Paneque}, {Perkins}, {Pesce-Rollins}, {Piron}, {Pivato},
  {Porter}, {Rain{\`o}}, {Rando}, {Razzano}, {Reimer}, {Reimer}, {Ritz},
  {S{\'a}nchez-Conde}, {Schulz}, {Sehgal}, {Sgr{\`o}}, {Siskind}, {Spada},
  {Spandre}, {Spinelli}, {Strigari}, {Tajima}, {Takahashi}, {Thayer},
  {Tibaldo}, {Torres}, {Troja}, {Vianello}, {Werner}, {Winer}, {Wood}, {Wood},
  {Zaharijas}, {Zimmer}, \& {Fermi-LAT Collaboration}}]{2015PhRvL.115w1301A}
{Ackermann}, M., {Albert}, A., {Anderson}, B., {et~al.} 2015, Physical Review
  Letters, 115, 231301

\bibitem[{{Aleksi{\'c}} {et~al.}(2014){Aleksi{\'c}}, {Ansoldi}, {Antonelli},
  {Antoranz}, {Babic}, {Bangale}, {Barres de Almeida}, {Barrio}, {Becerra
  Gonz{\'a}lez}, {Bednarek}, {Berger}, {Bernardini}, {Biland}, {Blanch},
  {Bock}, {Bonnefoy}, {Bonnoli}, {Borracci}, {Bretz}, {Carmona}, {Carosi},
  {Carreto Fidalgo}, {Colin}, {Colombo}, {Contreras}, {Cortina}, {Covino}, {Da
  Vela}, {Dazzi}, {De Angelis}, {De Caneva}, {De Lotto}, {Delgado Mendez},
  {Doert}, {Dom{\'{\i}}nguez}, {Dominis Prester}, {Dorner}, {Doro}, {Einecke},
  {Eisenacher}, {Elsaesser}, {Farina}, {Ferenc}, {Fonseca}, {Font}, {Frantzen},
  {Fruck}, {Garc{\'{\i}}a L{\'o}pez}, {Garczarczyk}, {Garrido Terrats}, {Gaug},
  {Giavitto}, {Godinovi{\'c}}, {Gonz{\'a}lez Mu{\~n}oz}, {Gozzini}, {Hadasch},
  {Hayashida}, {Herrero}, {Hildebrand}, {Hose}, {Hrupec}, {Idec}, {Kadenius},
  {Kellermann}, {Kodani}, {Konno}, {Krause}, {Kubo}, {Kushida}, {La Barbera},
  {Lelas}, {Lewandowska}, {Lindfors}, {Lombardi}, {L{\'o}pez},
  {L{\'o}pez-Coto}, {L{\'o}pez-Oramas}, {Lorenz}, {Lozano}, {Makariev},
  {Mallot}, {Maneva}, {Mankuzhiyil}, {Mannheim}, {Maraschi}, {Marcote},
  {Mariotti}, {Mart{\'{\i}}nez}, {Mazin}, {Menzel}, {Meucci}, {Miranda},
  {Mirzoyan}, {Moralejo}, {Munar-Adrover}, {Nakajima}, {Niedzwiecki},
  {Nilsson}, {Nishijima}, {Nowak}, {Orito}, {Overkemping}, {Paiano},
  {Palatiello}, {Paneque}, {Paoletti}, {Paredes}, {Paredes-Fortuny}, {Partini},
  {Persic}, {Prada}, {Prada Moroni}, {Prandini}, {Preziuso}, {Puljak},
  {Reinthal}, {Rhode}, {Rib{\'o}}, {Rico}, {Rodriguez Garcia}, {R{\"u}gamer},
  {Saggion}, {Saito}, {Saito}, {Salvati}, {Satalecka}, {Scalzotto}, {Scapin},
  {Schultz}, {Schweizer}, {Sillanp{\"a}{\"a}}, {Sitarek}, {Snidaric},
  {Sobczynska}, {Spanier}, {Stamatescu}, {Stamerra}, {Steinbring}, {Storz},
  {Sun}, {Suri{\'c}}, {Takalo}, {Takami}, {Tavecchio}, {Temnikov},
  {Terzi{\'c}}, {Tescaro}, {Teshima}, {Thaele}, {Tibolla}, {Torres}, {Toyama},
  {Treves}, {Uellenbeck}, {Vogler}, {Wagner}, {Zandanel}, {Zanin}, \&
  {Ibarra}}]{2014JCAP...02..008A}
{Aleksi{\'c}}, J., {Ansoldi}, S., {Antonelli}, L.~A., {et~al.} 2014, \jcap, 2,
  008

\bibitem[{{Aliu} {et~al.}(2012){Aliu}, {Archambault}, {Arlen}, {Aune},
  {Beilicke}, {Benbow}, {Bouvier}, {Bradbury}, {Buckley}, {Bugaev}, {Byrum},
  {Cannon}, {Cesarini}, {Christiansen}, {Ciupik}, {Collins-Hughes}, {Connolly},
  {Cui}, {Decerprit}, {Dickherber}, {Dumm}, {Errando}, {Falcone}, {Feng},
  {Ferrer}, {Finley}, {Finnegan}, {Fortson}, {Furniss}, {Galante}, {Gall},
  {Godambe}, {Griffin}, {Grube}, {Gyuk}, {Hanna}, {Holder}, {Huan}, {Hughes},
  {Humensky}, {Kaaret}, {Karlsson}, {Kertzman}, {Khassen}, {Kieda},
  {Krawczynski}, {Krennrich}, {Lee}, {Madhavan}, {Maier}, {Majumdar},
  {McArthur}, {McCann}, {Moriarty}, {Mukherjee}, {Ong}, {Orr}, {Otte}, {Park},
  {Perkins}, {Pohl}, {Prokoph}, {Quinn}, {Ragan}, {Reyes}, {Reynolds},
  {Roache}, {Rose}, {Ruppel}, {Saxon}, {Schroedter}, {Sembroski}, {{\c
  S}ent{\"u}rk}, {Skole}, {Smith}, {Staszak}, {Telezhinsky}, {Te{\v s}i{\'c}},
  {Theiling}, {Thibadeau}, {Tsurusaki}, {Varlotta}, {Vassiliev}, {Vincent},
  {Vivier}, {Wagner}, {Wakely}, {Ward}, {Weekes}, {Weinstein}, {Weisgarber},
  {Williams}, \& {Zitzer}}]{2012PhRvD..85f2001A}
{Aliu}, E., {Archambault}, S., {Arlen}, T., {et~al.} 2012, \prd, 85, 062001

\bibitem[{Aliu {et~al.}(2015)Aliu, Archambault, Arlen, Aune, Beilicke, Benbow,
  Bouvier, Bradbury, Buckley, Bugaev, Byrum, Cannon, Cesarini, Christiansen,
  Ciupik, Collins-Hughes, Connolly, Cui, Decerprit, Dickherber, Dumm, Errando,
  Falcone, Feng, Ferrer, Finley, Finnegan, Fortson, Furniss, Galante, Gall,
  Godambe, Griffin, Grube, Gyuk, Hanna, Holder, Huan, Hughes, Humensky, Kaaret,
  Karlsson, Kertzman, Khassen, Kieda, Krawczynski, Krennrich, Lee, Madhavan,
  Maier, Majumdar, McArthur, McCann, Moriarty, Mukherjee, Ong, Orr, Otte, Park,
  Perkins, Pohl, Prokoph, Quinn, Ragan, Reyes, Reynolds, Roache, Rose, Ruppel,
  Saxon, Schroedter, Sembroski, \ifmmode~\mbox{\c{S}}\else \c{S}\fi{}ent\"urk,
  Skole, Smith, Staszak, Telezhinsky, Te\ifmmode \check{s}\else
  \v{s}\fi{}i\ifmmode~\acute{c}\else \'{c}\fi{}, Theiling, Thibadeau,
  Tsurusaki, Varlotta, Vassiliev, Vincent, Vivier, Wagner, Wakely, Ward,
  Weekes, Weinstein, Weisgarber, Williams, \& Zitzer}]{PhysRevD.91.129903}
Aliu, E., Archambault, S., Arlen, T., {et~al.} 2015, Phys. Rev. D, 91, 129903

\bibitem[{{B.~Zitzer for the VERITAS
  Collaboration}(2015)}]{2015arXiv150300743B}
{B.~Zitzer for the VERITAS Collaboration}. 2015, ArXiv e-prints,
  arXiv:1503.00743

\bibitem[{{Battaglia} {et~al.}(2013){Battaglia}, {Helmi}, \&
  {Breddels}}]{2013NewAR..57...52B}
{Battaglia}, G., {Helmi}, A., \& {Breddels}, M. 2013, \nar, 57, 52

\bibitem[{{Berge} {et~al.}(2007){Berge}, {Funk}, \&
  {Hinton}}]{2007A&A...466.1219B}
{Berge}, D., {Funk}, S., \& {Hinton}, J. 2007, \aap, 466, 1219

\bibitem[{{Binney} \& {Tremaine}(2008)}]{2008gady.book.....B}
{Binney}, J., \& {Tremaine}, S. 2008, {Galactic Dynamics: Second Edition}
  (Princeton University Press)

\bibitem[{{Bonnivard} {et~al.}(2015){Bonnivard}, {Combet}, {Maurin}, \&
  {Walker}}]{2015MNRAS.446.3002B}
{Bonnivard}, V., {Combet}, C., {Maurin}, D., \& {Walker}, M.~G. 2015, \mnras,
  446, 3002

\bibitem[{{Bonnivard} {et~al.}(2016){Bonnivard}, {Maurin}, \&
  {Walker}}]{2016MNRAS.462..223B}
{Bonnivard}, V., {Maurin}, D., \& {Walker}, M.~G. 2016, \mnras, 462, 223

\bibitem[{{Chiu}(1966)}]{1966PhRvL..17..712C}
{Chiu}, H.-Y. 1966, Physical Review Letters, 17, 712

\bibitem[{{Christiansen} \& {VERITAS
  Collaboration}(2012)}]{2012AIPC.1505..709C}
{Christiansen}, J., \& {VERITAS Collaboration}. 2012, in American Institute of
  Physics Conference Series, Vol. 1505, American Institute of Physics
  Conference Series, ed. F.~A. {Aharonian}, W.~{Hofmann}, \& F.~M. {Rieger},
  709--712

\bibitem[{{Cirelli} {et~al.}(2011){Cirelli}, {Corcella}, {Hektor}, {H{\"u}tsi},
  {Kadastik}, {Panci}, {Raidal}, {Sala}, \& {Strumia}}]{2011JCAP...03..051C}
{Cirelli}, M., {Corcella}, G., {Hektor}, A., {et~al.} 2011, \jcap, 3, 51

\bibitem[{{D.~B.~Kieda for the VERITAS
  Collaboration}(2013)}]{2013arXiv1308.4849D}
{D.~B.~Kieda for the VERITAS Collaboration}. 2013, ArXiv e-prints

\bibitem[{{Fomin} {et~al.}(1994){Fomin}, {Stepanian}, {Lamb}, {Lewis}, {Punch},
  \& {Weekes}}]{1994APh.....2..137F}
{Fomin}, V.~P., {Stepanian}, A.~A., {Lamb}, R.~C., {et~al.} 1994, Astroparticle
  Physics, 2, 137

\bibitem[{{Geringer-Sameth} {et~al.}(2015{\natexlab{a}}){Geringer-Sameth},
  {Koushiappas}, \& {Walker}}]{2015ApJ...801...74G}
{Geringer-Sameth}, A., {Koushiappas}, S.~M., \& {Walker}, M.
  2015{\natexlab{a}}, \apj, 801, 74

\bibitem[{{Geringer-Sameth} {et~al.}(2015{\natexlab{b}}){Geringer-Sameth},
  {Koushiappas}, \& {Walker}}]{2015PhRvD..91h3535G}
{Geringer-Sameth}, A., {Koushiappas}, S.~M., \& {Walker}, M.~G.
  2015{\natexlab{b}}, \prd, 91, 083535

\bibitem[{{Hillas}(1985)}]{1985ICRC....3..445H}
{Hillas}, A.~M. 1985, International Cosmic Ray Conference, 3, 445

\bibitem[{{Holder} {et~al.}(2006){Holder}, {Atkins}, {Badran}, {Blaylock},
  {Bradbury}, {Buckley}, {Byrum}, {Carter-Lewis}, {Celik}, {Chow}, {Cogan},
  {Cui}, {Daniel}, {de La Calle Perez}, {Dowdall}, {Dowkontt}, {Duke},
  {Falcone}, {Fegan}, {Finley}, {Fortin}, {Fortson}, {Gibbs}, {Gillanders},
  {Glidewell}, {Grube}, {Gutierrez}, {Gyuk}, {Hall}, {Hanna}, {Hays}, {Horan},
  {Hughes}, {Humensky}, {Imran}, {Jung}, {Kaaret}, {Kenny}, {Kieda}, {Kildea},
  {Knapp}, {Krawczynski}, {Krennrich}, {Lang}, {Lebohec}, {Linton}, {Little},
  {Maier}, {Manseri}, {Milovanovic}, {Moriarty}, {Mukherjee}, {Ogden}, {Ong},
  {Petry}, {Perkins}, {Pizlo}, {Pohl}, {Quinn}, {Ragan}, {Reynolds}, {Roache},
  {Rose}, {Schroedter}, {Sembroski}, {Sleege}, {Steele}, {Swordy}, {Syson},
  {Toner}, {Valcarcel}, {Vassiliev}, {Wakely}, {Weekes}, {White}, {Williams},
  \& {Wagner}}]{2006APh....25..391H}
{Holder}, J., {Atkins}, R.~W., {Badran}, H.~M., {et~al.} 2006, Astroparticle
  Physics, 25, 391

\bibitem[{{Ichikawa} {et~al.}(2016){Ichikawa}, {Ishigaki}, {Matsumoto}, {Ibe},
  {Sugai}, \& {Hayashi}}]{2016arXiv160801749I}
{Ichikawa}, K., {Ishigaki}, M.~N., {Matsumoto}, S., {et~al.} 2016, ArXiv
  e-prints, arXiv:1608.01749

\bibitem[{{Jungman} {et~al.}(1996){Jungman}, {Kamionkowski}, \&
  {Griest}}]{1996PhR...267..195J}
{Jungman}, G., {Kamionkowski}, M., \& {Griest}, K. 1996, \physrep, 267, 195

\bibitem[{{Junk}(1999)}]{1999NIMPA.434..435J}
{Junk}, T. 1999, Nuclear Instruments and Methods in Physics Research A, 434,
  435

\bibitem[{{Krause} {et~al.}(2017){Krause}, {Pueschel}, \&
  {Maier}}]{2017APh....89....1K}
{Krause}, M., {Pueschel}, E., \& {Maier}, G. 2017, Astroparticle Physics, 89, 1

\bibitem[{{Krawczynski} {et~al.}(2006){Krawczynski}, {Carter-Lewis}, {Duke},
  {Holder}, {Maier}, {Le Bohec}, \& {Sembroski}}]{2006APh....25..380K}
{Krawczynski}, H., {Carter-Lewis}, D.~A., {Duke}, C., {et~al.} 2006,
  Astroparticle Physics, 25, 380

\bibitem[{{Li} \& {Ma}(1983)}]{1983ApJ...272..317L}
{Li}, T.-P., \& {Ma}, Y.-Q. 1983, \apj, 272, 317

\bibitem[{{MAGIC collaboration}(2016)}]{1475-7516-2016-02-039}
{MAGIC collaboration}. 2016, Journal of Cosmology and Astroparticle Physics,
  2016, 039

\bibitem[{{Mateo}(1998)}]{1998ARA&A..36..435M}
{Mateo}, M.~L. 1998, \araa, 36, 435

\bibitem[{{Mohanty} {et~al.}(1998){Mohanty}, {Biller}, {Carter-Lewis}, {Fegan},
  {Hillas}, {Lamb}, {Weekes}, {West}, \& {Zweerink}}]{1998APh.....9...15M}
{Mohanty}, G., {Biller}, S., {Carter-Lewis}, D.~A., {et~al.} 1998,
  Astroparticle Physics, 9, 15

\bibitem[{{Perkins} {et~al.}(2009){Perkins}, {Maier}, \& {The VERITAS
  Collaboration}}]{2009arXiv0912.3841P}
{Perkins}, J.~S., {Maier}, G., \& {The VERITAS Collaboration}. 2009, ArXiv
  e-prints

\bibitem[{{Planck Collaboration} {et~al.}(2014){Planck Collaboration}, {Ade},
  {Aghanim}, {Armitage-Caplan}, {Arnaud}, {Ashdown}, {Atrio-Barandela},
  {Aumont}, {Baccigalupi}, {Banday}, {Barreiro}, {Bartlett}, {Battaner},
  {Benabed}, {Beno{\^\i}t}, {Benoit-L{\'e}vy}, {Bernard}, {Bersanelli},
  {Bielewicz}, {Bobin}, {Bock}, {Bonaldi}, {Bond}, {Borrill}, {Bouchet},
  {Bridges}, {Bucher}, {Burigana}, {Butler}, {Calabrese}, {Cappellini},
  {Cardoso}, {Catalano}, {Challinor}, {Chamballu}, {Chary}, {Chen}, {Chiang},
  {Chiang}, {Christensen}, {Church}, {Clements}, {Colombi}, {Colombo},
  {Couchot}, {Coulais}, {Crill}, {Curto}, {Cuttaia}, {Danese}, {Davies},
  {Davis}, {de Bernardis}, {de Rosa}, {de Zotti}, {Delabrouille}, {Delouis},
  {D{\'e}sert}, {Dickinson}, {Diego}, {Dolag}, {Dole}, {Donzelli}, {Dor{\'e}},
  {Douspis}, {Dunkley}, {Dupac}, {Efstathiou}, {Elsner}, {En{\ss}lin},
  {Eriksen}, {Finelli}, {Forni}, {Frailis}, {Fraisse}, {Franceschi}, {Gaier},
  {Galeotta}, {Galli}, {Ganga}, {Giard}, {Giardino}, {Giraud-H{\'e}raud},
  {Gjerl{\o}w}, {Gonz{\'a}lez-Nuevo}, {G{\'o}rski}, {Gratton}, {Gregorio},
  {Gruppuso}, {Gudmundsson}, {Haissinski}, {Hamann}, {Hansen}, {Hanson},
  {Harrison}, {Henrot-Versill{\'e}}, {Hern{\'a}ndez-Monteagudo}, {Herranz},
  {Hildebrandt}, {Hivon}, {Hobson}, {Holmes}, {Hornstrup}, {Hou}, {Hovest},
  {Huffenberger}, {Jaffe}, {Jaffe}, {Jewell}, {Jones}, {Juvela},
  {Keih{\"a}nen}, {Keskitalo}, {Kisner}, {Kneissl}, {Knoche}, {Knox}, {Kunz},
  {Kurki-Suonio}, {Lagache}, {L{\"a}hteenm{\"a}ki}, {Lamarre}, {Lasenby},
  {Lattanzi}, {Laureijs}, {Lawrence}, {Leach}, {Leahy}, {Leonardi},
  {Le{\'o}n-Tavares}, {Lesgourgues}, {Lewis}, {Liguori}, {Lilje},
  {Linden-V{\o}rnle}, {L{\'o}pez-Caniego}, {Lubin}, {Mac{\'{\i}}as-P{\'e}rez},
  {Maffei}, {Maino}, {Mandolesi}, {Maris}, {Marshall}, {Martin},
  {Mart{\'{\i}}nez-Gonz{\'a}lez}, {Masi}, {Massardi}, {Matarrese}, {Matthai},
  {Mazzotta}, {Meinhold}, {Melchiorri}, {Melin}, {Mendes}, {Menegoni},
  {Mennella}, {Migliaccio}, {Millea}, {Mitra}, {Miville-Desch{\^e}nes},
  {Moneti}, {Montier}, {Morgante}, {Mortlock}, {Moss}, {Munshi}, {Murphy},
  {Naselsky}, {Nati}, {Natoli}, {Netterfield}, {N{\o}rgaard-Nielsen},
  {Noviello}, {Novikov}, {Novikov}, {O'Dwyer}, {Osborne}, {Oxborrow}, {Paci},
  {Pagano}, {Pajot}, {Paladini}, {Paoletti}, {Partridge}, {Pasian},
  {Patanchon}, {Pearson}, {Pearson}, {Peiris}, {Perdereau}, {Perotto},
  {Perrotta}, {Pettorino}, {Piacentini}, {Piat}, {Pierpaoli}, {Pietrobon},
  {Plaszczynski}, {Platania}, {Pointecouteau}, {Polenta}, {Ponthieu}, {Popa},
  {Poutanen}, {Pratt}, {Pr{\'e}zeau}, {Prunet}, {Puget}, {Rachen}, {Reach},
  {Rebolo}, {Reinecke}, {Remazeilles}, {Renault}, {Ricciardi}, {Riller},
  {Ristorcelli}, {Rocha}, {Rosset}, {Roudier}, {Rowan-Robinson},
  {Rubi{\~n}o-Mart{\'{\i}}n}, {Rusholme}, {Sandri}, {Santos}, {Savelainen},
  {Savini}, {Scott}, {Seiffert}, {Shellard}, {Spencer}, {Starck}, {Stolyarov},
  {Stompor}, {Sudiwala}, {Sunyaev}, {Sureau}, {Sutton}, {Suur-Uski}, {Sygnet},
  {Tauber}, {Tavagnacco}, {Terenzi}, {Toffolatti}, {Tomasi}, {Tristram},
  {Tucci}, {Tuovinen}, {T{\"u}rler}, {Umana}, {Valenziano}, {Valiviita}, {Van
  Tent}, {Vielva}, {Villa}, {Vittorio}, {Wade}, {Wandelt}, {Wehus}, {White},
  {White}, {Wilkinson}, {Yvon}, {Zacchei}, \& {Zonca}}]{2014A&A...571A..16P}
{Planck Collaboration}, {Ade}, P.~A.~R., {Aghanim}, N., {et~al.} 2014, \aap,
  571, A16

\bibitem[{{Read}(2002)}]{2002JPhG...28.2693R}
{Read}, A.~L. 2002, Journal of Physics G Nuclear Physics, 28, 2693

\bibitem[{{Rolke} {et~al.}(2005){Rolke}, {L{\'o}pez}, \&
  {Conrad}}]{2005NIMPA.551..493R}
{Rolke}, W.~A., {L{\'o}pez}, A.~M., \& {Conrad}, J. 2005, Nuclear Instruments
  and Methods in Physics Research A, 551, 493

\bibitem[{{Rowell}(2003)}]{2003A&A...410..389R}
{Rowell}, G.~P. 2003, \aap, 410, 389

\bibitem[{{Servant} \& {Tait}(2003)}]{2003NuPhB.650..391S}
{Servant}, G., \& {Tait}, T.~M.~P. 2003, Nuclear Physics B, 650, 391

\bibitem[{{Simon} \& {Geha}(2007)}]{simon07}
{Simon}, J.~D., \& {Geha}, M. 2007, \apj, 670, 313

\bibitem[{{Steigman}(1979)}]{1979ARNPS..29..313S}
{Steigman}, G. 1979, Annual Review of Nuclear and Particle Science, 29, 313

\bibitem[{{Steigman} {et~al.}(2012){Steigman}, {Dasgupta}, \&
  {Beacom}}]{2012PhRvD..86b3506S}
{Steigman}, G., {Dasgupta}, B., \& {Beacom}, J.~F. 2012, \prd, 86, 023506

\bibitem[{{Strigari}(2013)}]{2013PhR...531....1S}
{Strigari}, L.~E. 2013, \physrep, 531, 1

\bibitem[{{Strigari} {et~al.}(2007){Strigari}, {Koushiappas}, {Bullock}, \&
  {Kaplinghat}}]{2007PhRvD..75h3526S}
{Strigari}, L.~E., {Koushiappas}, S.~M., {Bullock}, J.~S., \& {Kaplinghat}, M.
  2007, \prd, 75, 083526

\bibitem[{{Strigari} {et~al.}(2008){Strigari}, {Koushiappas}, {Bullock},
  {Kaplinghat}, {Simon}, {Geha}, \& {Willman}}]{2008ApJ...678..614S}
{Strigari}, L.~E., {Koushiappas}, S.~M., {Bullock}, J.~S., {et~al.} 2008, \apj,
  678, 614

\bibitem[{{Walker}(2013)}]{2013pss5.book.1039W}
{Walker}, M. 2013, {Dark Matter in the Galactic Dwarf Spheroidal Satellites},
  ed. T.~D. {Oswalt} \& G.~{Gilmore}, 1039

\bibitem[{{Walker, Mateo \& Olszewski}(2009)}]{walker09a}
{Walker, Mateo \& Olszewski}. 2009, \aj, 137, 3100

\bibitem[{{Willman} {et~al.}(2011){Willman}, {Geha}, {Strader}, {Strigari},
  {Simon}, {Kirby}, {Ho}, \& {Warres}}]{2011AJ....142..128W}
{Willman}, B., {Geha}, M., {Strader}, J., {et~al.} 2011, \aj, 142, 128

\bibitem[{{Zeldovic} {et~al.}(1965){Zeldovic}, {Okun}, \&
  {Pikelner}}]{1965PhL....17..164Z}
{Zeldovic}, Y.~B., {Okun}, L.~B., \& {Pikelner}, S.~B. 1965, Physics Letters,
  17, 164

\bibitem[{Zel'dovich(1965)}]{zel1965advances}
Zel'dovich, Y.~B. 1965, Advances in Astronomy and Astrophysics, ed. Z.~Kopal,
  Vol.~3 (Academic Press), 242

\bibitem[{{Zhao}(1996)}]{1996MNRAS.278..488Z}
{Zhao}, H. 1996, \mnras, 278, 488

\bibitem[{{Zhu} {et~al.}(2016){Zhu}, {van de Ven}, {Watkins}, \&
  {Posti}}]{2016MNRAS.463.1117Z}
{Zhu}, L., {van de Ven}, G., {Watkins}, L.~L., \& {Posti}, L. 2016, \mnras,
  463, 1117

\bibitem[{{Zitzer} \& {for the VERITAS
  Collaboration}(2013{\natexlab{a}})}]{2013arXiv1307.8367Z}
{Zitzer}, B., \& {for the VERITAS Collaboration}. 2013{\natexlab{a}}, ArXiv
  e-prints, arXiv:1307.8367

\bibitem[{{Zitzer} \& {for the VERITAS
  Collaboration}(2013{\natexlab{b}})}]{2013arXiv1307.8360Z}
---. 2013{\natexlab{b}}, ArXiv e-prints, arXiv:1307.8360

\end{thebibliography}

\end{document}